\documentclass[twocolumn,amsmath,floatfix]{revtex4}
\usepackage{latexsym}
\usepackage{amssymb}
\usepackage{pdfpages}
\usepackage{graphics}

\begin{document}
\newcommand{\ja}{Jakuba\ss a-Amundsen }
\newcommand{\bfx}{\mbox{\boldmath $x$}}
\newcommand{\bfq}{\mbox{\boldmath $q$}}
\newcommand{\bfnabla}{\mbox{\boldmath $\nabla$}}
\newcommand{\bfalpha}{\mbox{\boldmath $\alpha$}}
\newcommand{\bfsigma}{\mbox{\boldmath $\sigma$}}
\newcommand{\bfeps}{\mbox{\boldmath $\epsilon$}}
\newcommand{\bfA}{\mbox{\boldmath $A$}}
\newcommand{\bfP}{\mbox{\boldmath $P$}}
\newcommand{\bfe}{\mbox{\boldmath $e$}}
\newcommand{\bfn}{\mbox{\boldmath $n$}}
\newcommand{\bfW}{{\mbox{\boldmath $W$}_{\!\!rad}}}
\newcommand{\bfM}{\mbox{\boldmath $M$}}
\newcommand{\bfI}{\mbox{\boldmath $I$}}
\newcommand{\bfQ}{\mbox{\boldmath $Q$}}
\newcommand{\bfp}{\mbox{\boldmath $p$}}
\newcommand{\bfk}{\mbox{\boldmath $k$}}
\newcommand{\bfks}{\mbox{{\scriptsize \boldmath $k$}}}
\newcommand{\bfqs}{\mbox{{\scriptsize \boldmath $q$}}}
\newcommand{\bfxs}{\mbox{{\scriptsize \boldmath $x$}}}
\newcommand{\bfs}{\mbox{\boldmath $s$}_0}
\newcommand{\bfv}{\mbox{\boldmath $v$}}
\newcommand{\bfw}{\mbox{\boldmath $w$}}
\newcommand{\bfb}{\mbox{\boldmath $b$}}
\newcommand{\bfxi}{\mbox{\boldmath $\xi$}}
\newcommand{\bfzeta}{\mbox{\boldmath $\zeta$}}
\newcommand{\bfr}{\mbox{\boldmath $r$}}
\newcommand{\bfrs}{\mbox{{\scriptsize \boldmath $r$}}}

\renewcommand{\theequation}{\arabic{section}.\arabic{equation}}
\renewcommand{\thesection}{\arabic{section}}
\renewcommand{\thesubsection}{\arabic{section}.\arabic{subsection}}

\title{\Large\bf On the QED corrections to elastic electron scattering\\ at high momentum transfer}

\author{D.~H.~Jakubassa-Amundsen \\
Mathematics Institute, University of Munich, Theresienstrasse 39,\\ 80333 Munich, Germany}

%\date{}
\date{\today}
%\maketitle

\vspace{1cm}

\begin{abstract}  
Estimates of QED and dispersion effects on the cross section for elastic electron scattering from a $^{12}$C nucleus are provided for collision energies in the range of $120-450$ MeV.
While in general such corrections are smoothly varying with energy or scattering angle,
they show structures in the vicinity of diffraction minima which are very sensitive to details of the theoretical models.
This casts doubt on the assertion that the discrepancy between QED background-corrected experimental data and theory in these minima originates solely from dispersion.
\end{abstract}

\maketitle

\section{Introduction}

In order to determine nuclear charge distributions or nuclear charge radii from the measurements  of elastic electron scattering  by means of a comparison with phase-shift or distorted-wave Born calculations \cite{FW66}, the experi\-mental  data are corrected for quantum electrodynamical (QED) effects.
For such nuclear structure investigations light targets ranging from protons to carbon are used, where Coulomb distortion effects are assumed to be small.

Usually these QED corrections rely on the plane-wave Born approximation (PWBA) for the vacuum polarization, the vertex and self-energy correction and the radiation of soft unobserved photons.
For high collision energies, simple formulae are available to account for these effects (see, e.g. \cite{Tsa61,Ma69,MT00,BS19}).

In the past decades great efforts were made to understand the deviation of the  so corrected experimental scattering cross section data from theoretical predictions in the region of diffraction,
arising from the charge distribution inside the target nucleus \cite{Ub}.
It is well-known, however, that the PWBA,  an appropriate high-energy theory for weak fields at low momentum transfer, fails to correctly describe the scattering process
when diffraction effects 
 modulate the electron intensity.
The displacement between the calculated PWBA position of the cross section minima and experiment is conventionally handled by introducing an effective momentum transfer $q_{\rm eff}$ at which the PWBA theory
has to be evaluated \cite{DS84}, but substantial deviations in intensity remain.

The consideration of the QED effects beyond  first order in the fine-structure constant $\alpha$ is non-trivial.
Second-order radiative corrections pertaining to the electron were recently reported for proton targets  \cite{AK15,BS19}.
The inclusion of higher-order interactions with the target potential is more involved, and is only straightforward in case of the vacuum polarization by means of the Uehling potential \cite{Ueh,Kla77}.

The largest QED effect, the vertex and self-energy correction plus the contribution from soft bremsstrahlung (in the following termed vsb correction), may amount up to 20\% in first order for collision energies around 100 MeV   \cite{Ma69}.
A calculation of the second-order (in $Z\alpha$,
where $Z$ is the nuclear charge number) vsb effects was attempted in \cite{MU} and further literature is provided in the review by Maximon \cite{Ma69}, but no tractable formula is available.
A full account of the electron-target field for this effect has  not yet been accomplished.

The assumption why such higher-order corrections can be disregarded
in the data reduction is based on the fact that the 
first-order Born amplitude, which multiplies the 
QED corrections, vanishes in the diffraction minima.
Hence the remaining discrepancies between experiment
and theory are attributed to dispersion, a second-order Born contribution to Coulomb scattering which allows for an intermediate excitation of the nucleus \cite{L56,FW66,FR74,HR98,GH08}.
 Such deviations are of the order of $5-10\%$ at collision energies between $300-500$ MeV (increasing with energy \cite{Jef20}).
However, estimates for the dispersion  are commonly in the percent region \cite{FR74} and cannot account for these discrepancies.

The present work demonstrates the sensitivity of the QED corrections to the choice of different theoretical prescriptions.
In particular, the PWBA result is set against an improved model where Coulomb distortion is accounted for by replacing the PWBA potential scattering amplitudes by the phase-shift results in all next-to-leading order terms in $\alpha$, as suggested in \cite{Ma69}.

The $^{12}$C nucleus is chosen as target because
of its importance as a reference nucleus for nuclear structure studies \cite{Reu82}, 
but it will also be used in future experiments on parity violation \cite{Au11}. Its central field is weak enough so that the PWBA is valid at small angles.
Moreover, it is a spin-zero nucleus, for which any scattering from magnetic moment distributions is absent.
Being a $p$-shell nucleus, $^{12}$C can  reasonably  well be described within the harmonic oscillator shell model \cite{FR74}.
Even realistic many-body interactions can in principle be taken into account for this nucleus \cite{Lo16}.
Finally, there exists a large number of high-precision elastic electron scattering measurements \cite{Reu82,Off91,Jef20,Ka89}. 

The paper is organized as follows. Section 2 recapitulates the QED corrections within the PWBA in the high-energy approximation. Dispersion is considered using the second-order Born theory with a closure approximation. The so corrected differential cross section is provided both in the Born approximation and when Coulomb distortion is taken into account by means of the phase-shift analysis.
In section 3, numerical results for the  differential cross sections and their change with QED effects and dispersion are provided.
Comparison is made with experimental data in the energy region $150-430$ MeV.
Concluding remarks are given in section 4. Atomic units ($\hbar=m=e=1$) are used unless indicated otherwise.

\section{QED and dispersion corrections}

We start by providing the QED corrections to elastic electron scattering from spin-zero nuclei, consisting of the vacuum polarization and the vertex, self-energy and soft bremsstrahlung (vsb) contributions.
Since the experimental data under consideration are recorded with a high-resolution spectrometer, hard bremsstrahlung, where the photon momentum has to be fully taken into account, does not contribute.

\subsection{Vacuum polarization}

In lowest-order Born approximation, the transition amplitude for vacuum polarization is given by \cite{MT00}
\begin{equation}\label{2.1}
A_{fi}^{\rm vac}\;=\;
 \frac{1}{3 \pi c}\left[ \ln(-q^2/c^2)\,-\;\frac{5}{3}\right]\,A_{fi}^{B1},
\end{equation}
Here, $A_{fi}^{B1}$ is the first-order Born amplitude for elastic scattering. The 4-momentum transfer $q$ to the nucleus is defined by $q^2=(E_i-E_f)^2/c^2 -\bfq^2$ 
where $\bfq=\bfk_i-\bfk_f$, and $E_i,\bfk_i$, respectively $E_f,\bfk_f$ are the total energies and momenta of incoming and scattered electron.
The validity of (\ref{2.1}) is restricted to high momentum transfer, 
 $-q^2/c^2\gg 1$, which covers our cases of interest. 

Alternatively, vacuum polarization can be calculated to all orders in $Z/c$ with the help of the Uehling potential \cite{Ueh},
\begin{equation}\label{2.2}
U_e(r)\;=\;-\;\frac{2}{3\pi c} \int d\bfr'\;\frac{\varrho_N(\bfr')}{|\bfr-\bfr'|}\;\chi_1(2c|\bfr-\bfr'|),
\end{equation}
which for a spherical nuclear charge distribution $\varrho_N(r')$, normalized to $Z$, reduces to \cite{Kla77}
$$U_e(r)\;=\;-\;\frac{2}{3c^2r}\int_0^\infty r'dr'\;\varrho_N(r') $$
$$\times \left[ \chi_2(2c|r-r'|)\;-\;\chi_2(2c|r+r'|)\right],$$
\begin{equation}\label{2.3}
\chi_n(x)\;=\;\int_1^\infty dt\;e^{-xt}\;t^{-n}\left(1\,+\,\frac{1}{2t^2}\right)\left(1\,-\,\frac{1}{t^2}\right)^\frac12.
\end{equation}
For numerical estimates one can use a parametrization of $\chi_2$ in terms of rational functions of polynomials \cite{FR76}.

To include the effect of vacuum polarization in elastic scattering exactly,
the Uehling potential has to be added to the target nuclear potential $V_T(r)$ when performing the phase-shift analysis.
We have confirmed numerically that the first-order Born approximation to the Uehling potential agrees with (\ref{2.1}) within 0.01 \%, which validates the comparison of the exact result with the one based on (\ref{2.1}).

\subsection{Vertex correction and self energy}

The lowest-order Born amplitude for the vertex correction, after eliminating the UV divergence by renormalizing via the inclusion of the self energy, is given by \cite{BS19}
$$A_{fi}^{\rm vs}\;=\; F_1(-q^2)\;A_{fi}^{B1}$$
\begin{equation}\label{2.4}
-\;\frac{\sqrt{E_iE_f}}{c^3}\;\frac{Z}{\bfq^2}\;F_L(\bfq)\;F_2(-q^2)\;(u_{k_f}^{(\sigma_f)+} \gamma_0\;(\bfalpha \bfq)\;u_{k_i}^{(\sigma_i)}),
\end{equation}
where $\bfalpha$ and $\gamma_0$ denote Dirac matrices.
The initial, respectively final states of the electron (with spin polarization $\sigma_i$ and $\sigma_f$)
are represented by the free 4-spinors $u_{k_i}^{(\sigma_i)}$ and $u_{k_f}^{(\sigma_f)}$.
The longitudinal (Coulombic) form factor $F_L$ is calculated from 
the Fourier transform of the nuclear charge distribution $\varrho_N$,
\begin{equation}\label{2.5}
F_L(\bfq)=\;\frac{1}{Z}\int d\bfx_N\;\varrho_N(\bfx_N)\;e^{i\bfqs \bfxs_N}.
\end{equation}
 It is normalized to unity at $|\bfq|=0$.
In the high-energy approximation, $-q^2/c^2\gg 1$,
 the electric ($F_1$) and magnetic ($F_2)$ electron form factors are given by
$$F_1(-q^2)\;=\;\frac{1}{2\pi c}\left\{\frac12 \;\ln(-q^2/c^2)\;\left[ 3\,-\,\ln(-q^2/c^2)\right]\right.$$
\begin{equation}\label{2.6}
\left. -2\;+\;\frac{\pi^2}{6}\right]\;+\;\mbox{IR},
\end{equation}
\begin{equation}\label{2.7}
F_2(-q^2)\;=\;-\;\frac{1}{\pi c}\;\frac{c^2}{q^2}\;\ln(-q^2/c^2),
\end{equation}
and the infrared divergent term IR reads
\begin{equation}\label{2.8}
\mbox{IR}\;=\;\frac{1}{2\pi c}\;\left( \ln\,\lambda^2 \right)\left[ \ln(-q^2/c^2)\,-\,1\right],
\end{equation}
where $\lambda$ is an auxiliary finite photon mass.

\subsection{Soft bremsstrahlung} 

Since the electron detector has a finite energy resolution $\Delta E$, electrons which have lost an energy $\omega < \Delta E$ by means of soft photon emission cannot be distinguished from the elastically scattered electrons.
Therefore this soft-photon bremsstrahlung has to be added incoherently to the cross section for 
elastic scattering as long as the photons are not observed.
To lowest order  Born within the high-energy approximation, $-q^2/c^2 \gg 1$, the differential cross section for the soft photon emission is given by \cite{MT00}
$$\frac{d\sigma^{\rm soft}}{d\Omega_f}\;=\;\left[ -2\,\mbox{ IR}\,+\;\frac{1}{\pi c}\;\left\{\left(\ln(-q^2/c^2)\,-1\right)\;\ln\,\frac{\omega_0^2}{E_iE_f}\right.\right.$$
$$ +\,\frac12\left( \ln(-q^2/c^2)\right)^2\,-\,\frac12\left(\ln\,\frac{E_i}{E_f}\right)^2\,+\,\mbox{Li }\left(\cos^2\frac{\vartheta_f}{2}\right)$$
\begin{equation}\label{2.9}
 \left.\left.
-\;\frac{\pi^2}{3}\right\}\right]\;\left| A_{fi}^{B1}\right|^2,
\end{equation}
where $\vartheta_f$ is the scattering angle and $\omega_0$ is the upper limit of radiation.
 Li$(x)=-\int_0^x dt \frac{\ln|1-t|}{t}\;$ is the Spence function \cite{Tsa61}.
 In the derivation of this formula, the photon momentum $\bfk$ is partly neglected in the propagators before the integration over the photon degrees of freedom is carried out \cite{Tsa61,MT00}.

\subsection{Dispersion correction}

The dispersion correction is calculated from the box dia\-gram, which accounts for two virtual photon couplings between electron and nucleus.
The corresponding $S$-matrix element is given by \cite{BD,FR74}
$$S_{fi}^{\rm box}\;=\;-i\left( \frac{e}{c}\right)^2\int d^4x_e\;d^4y_e\;\bar{\psi}_f(y_e)\;\gamma_\mu$$
\begin{equation}\label{2.10}
\times \; S_F(y_e-x_e)\;\gamma_\nu\;\psi_i(x_e)\;A^{\mu \nu}(y_e,x_e),
\end{equation}
where $\psi_i$ and  $\psi_f$ are the plane-wave electronic scattering states, 
$\psi_n(x_e)=(2\pi)^{-2}e^{-ik_nx_e}u_{k_n}^{(\sigma_n)},\;n=i,f$.
$\gamma_\mu,\; \mu=0,...,3$ are Dirac matrices and
$S_F$ is the electron propagator, defined by
\begin{equation}\label{2.11}
S_F(y-x)\;=\;\int \frac{d^4p}{(2\pi)^4}\;e^{-ip(y-x)}\;\frac{cp_\mu \gamma^\mu +mc^2}{p^2-m^2c^2+i\epsilon},
\end{equation}
 and $A^{\mu \nu}$ is the photon field, which for the direct term is calculated from
$$A^{\mu \nu}(y_e,x_e)\;=\;4\pi i\,(Ze)^2\int d^4x_N\;d^4y_N \;D_0(y_e-y_N)$$
\begin{equation}\label{2.12}
\times \;D_0(x_e-x_N) \;\bar{\phi}_f(y_N)\;\gamma_0\;J^\mu 
\; S_N(y_N-x_N)\;\gamma_0\;J^\nu\;\phi_i(x_N),
\end{equation}
while in the exchange term, $x_N$ and $y_N$ are interchanged in the photon propagators $D_0$,
\begin{equation}\label{2.13}
D_0(x-y)\;=\;-\int \frac{d^4Q}{(2\pi)^4}\;\frac{e^{-iQ(x-y)}}{Q^2+i\epsilon},\qquad Q=(Q_0,\bfQ),
\end{equation}
 and $\mu$ and $\nu$ are interchanged in the operators $J^\mu$ and $J^\nu$ for the 4-currents of the nucleus.
The nuclear propagator is represented by
\begin{equation}\label{2.14}
S_N(y_N-x_N)\;=\;\sum_n \int \frac{d^4P_n}{(2\pi)^4}\;\frac{\phi_n(y_N)\,\bar{\phi}_n(x_N)}{P_{0n}-E_n/c\,+\,i\epsilon},
\end{equation}
where $P_n=(P_{0n},\bfP_n)$ is the 4-momentum of an intermediate nuclear state $\phi_n$ and $E_n=\sqrt{ \bfP_n^2c^2+M^2c^4}\,+\omega_n$ its total energy with $\omega_n$ the nuclear excitation energy, and $M$ is the target mass number.

We will restrict ourselves to a pure Coulombic excitation, $\mu=\nu=0$, 
while neglecting the magnetic interaction. In this approximation, one has
$$\bar{\phi}_n(x_N)\;\gamma_0\;J^\nu\;\phi_i(x_N)\;=\;e^{i(P_n-P_i)x_N}$$
\begin{equation}\label{2.15}
\times\; \langle n\,|\,J^0(\bfP_n-\bfP_i)\,|\,i\rangle\;\delta_{\nu,0},
\end{equation}
where $|\,n\rangle$ comprises the internal quantum numbers of the nuclear state $\phi_n$.
The evaluation of the sum (\ref{2.14}) over a complete set of nuclear states by means of the Greens function method is very involved \cite{Lo16}. Therefore, following Friar and Rosen \cite{FR74},
 we resort to the closure approximation by fixing $\omega_n =\bar{\omega}$ with an appropriate choice $\bar{\omega}=15$ MeV, which represents the mean excitation energy of the giant dipole resonance.
Friar and Rosen suggested some additional approximations for the evaluation of the $S$-matrix element. 
We adopt the following ones:  the exchange term to (\ref{2.12}) is dropped,  the energy difference $Q_0$ in the denominator of the photon propagators is omitted, and
the contribution from the pole of the electron propagator is neglected. With these approximations, the transition amplitude for the box diagram, 
defined by
\begin{equation}\label{2.16}
S_{fi}^{\rm box}\;=\;-i\;\delta(P_f-P_i+k_f-k_i)\;M_{fi}^{\rm box},
\end{equation}
 is given by 
$$M_{fi}^{\rm box}=\left( \frac{Ze^2}{c}\right)^2\frac{4\pi \,c}{(2\pi)^3}\int \frac{d\bfq_1}{\bfq_1^2}\;\frac{1}{(\bfq-\bfq_1)^2}\;C^{00}(\bfq_1,\bfq-\bfq_1)$$
\begin{equation}\label{2.17}
\times\frac{( u_{k_f}^{(\sigma_f)+}[ E_i-\bar{\omega}-\frac{\bfqs_1^2}{2M}+c\bfalpha(\bfk_i-\bfq_1)+mc^2\gamma_0]u_{k_i}^{(\sigma_i)})}{\left( E_i-\bar{\omega}-\frac{\bfqs_1^2}{2M}\right)^2 -(\bfk_i-\bfq_1)^2c^2-m^2c^4+i\epsilon}.
\end{equation}

The dispersion correction comprises only inelastic intermediate nuclear states, i.e. $n \neq i$ in (\ref{2.14}), whereas the elastic contribution ($n = i$) can be included by replacing
the Born amplitude $A_{fi}^{B1}$ in the leading term with the
respective phase-shift result.
Since elastic scattering implies $|i\rangle = |f\rangle =|0\rangle$, where $|0\rangle$ denotes the nuclear
ground state, the elastic contribution is characterized by
\begin{equation}\label{2.18}
\langle 0\,|\,J^0(\bfq)\,|\,0\rangle\;=\;Z\, F_L(\bfq).
\end{equation}
 Subtracting this term,
the resulting $A_{fi}^{\rm box}$ is identified with the dispersion amplitude, and the respective
 correlation function $C^{00}$ in (\ref{2.17}) is, using closure,  given by
\begin{equation}\label{2.19}
C^{00}(\bfq_1,\bfq_2)=\frac{1}{Z^2}\;\langle 0\,|\,J^0(\bfq_1)\,J^0(\bfq_2)\,|\,0\,\rangle
-\,F_L(\bfq_1)\,F_L(\bfq_2).
\end{equation}
Within the harmonic oscillator model for $^{12}$C, the correlation function can be reduced to an expression containing $F_L$ and the longitudinal proton form factor. It is explicitly provided in \cite{FR74}.

\subsection{Differential cross  section}

The leading term in the PWBA transition amplitude for elastic scattering is given by
\begin{equation}\label{2.20}
A_{fi}^{B1} \;=\;-\;\frac{1}{c} \; \frac{2Z\sqrt{E_iE_f}}{\bfq^2 c}\;\left( u_{k_f}^{(\sigma_f)+}\;u_{k_i}^{(\sigma_i)}\right)\;F_L(\bfq).
\end{equation}
In this expression, the magnetic current-current interaction is omitted, since $^{12}$C is a spin-zero nucleus. 
Correspondingly, the  factor $(-q^2$) in the denominator of the general theory has been replaced by $\bfq^2$ \cite{Gro}.
This is the same approximation as applied to the photon propagators in the box diagram.

The corresponding differential cross section reads
\begin{equation}\label{2.21}
\frac{d\sigma^{B1}}{d\Omega_f}\;=\;\frac{|\bfk_f|}{|\bfk_i|}\;\frac{1}{f_{\rm rec}}\;\frac12\sum_{\sigma_i \sigma_f}\left| A_{fi}^{B1}\right|^2.
\end{equation}
The cross section is reduced by the recoil factor $f_{\rm rec}$ because of the finite momentum $\bfq$ of the recoiling nucleus \cite{DS84},
\begin{equation}\label{2.22}
f_{\rm rec}\;=\;1\;-\;\frac{q^2E_f}{2Mc^2\bfk_f^2}\;\left(1\;-\;\frac{c^2}{ME_f}\right).
\end{equation}
Therefore, $E_f$ is strictly less than $E_i$.
Since spin polarization of the electron is not considered, an average over $\sigma_i$ and a sum over $\sigma_f$ has to be included in (\ref{2.21}).

The  total cross section, accounting for the  QED corrections and for the second-order elastic $(A_{fi}^{B2})$ and inelastic $(A_{fi}^{\rm box})$ amplitudes,
is in PWBA calculated from \cite{Lan}
$$\frac{d\sigma^{\rm Born}}{d\Omega_f}\;=\;\frac{|\bfk_f|}{|\bfk_i|}\;\frac{1}{f_{\rm rec}}\;\frac12\;\sum_{\sigma_i\sigma_f}\left[ \left| A_{fi}^{B1}\right|^2 \right.$$
\begin{equation}\label{2.23}
\left.+ \;2\mbox{ Re}\left\{ A_{fi}^{\ast B_1}\left( A_{fi}^{\rm vac}+ A_{fi}^{\rm vs} + A_{fi}^{B2} + A_{fi}^{\rm box}\right) \right\}
\,+\,\frac{d\sigma^{\rm soft}}{d\Omega_f}\right],
\end{equation}
$$A_{fi}^{\rm box}\;=\;2\;\frac{\sqrt{E_iE_f}}{c}\;M_{fi}^{\rm box},$$
such that the IR terms in (\ref{2.6}) and (\ref{2.9}) cancel.
Recalling that $A_{fi}^{B1} \sim Z\alpha$ (with $\alpha =\frac{1}{c}$),  the terms in (\ref{2.23}) proportional to $A_{fi}^{\rm vac}$ and $A_{fi}^{\rm vs}$ are of order $Z^2\alpha^3$ as is $\frac{d\sigma^{\rm soft}}{d\Omega_f}$, while those relating to $A_{fi}^{B2}$ and $A_{fi}^{\rm box}$ are of order $Z^3\alpha^3$. Hence (\ref{2.23}) is consistent to third order in $\alpha$.
It should be noted that quadratic terms like $|A_{fi}^{\rm box}|^2$ have to be omitted in the total cross section, since they are of higher than third order in $\alpha$.
Occasionally it is argued that $|A_{fi}^{\rm box}|^2$
 is important in the diffraction minima where $A_{fi}^{B1}$ is zero, and is therefore retained \cite{FR74,Jef20}.
However, there are further contributions of the same order $(\alpha^4$), like $|A_{fi}^{B2}|^2$ or $(A_{fi}^{\ast B2}+A_{fi}^{\ast {\rm box}}) (A_{fi}^{\rm vac} +A_{fi}^{\rm vs})$, which contribute in the diffraction minimum and hence should be considered if $|A_{fi}^{\rm box}|^2$ is.

In order to account for Coulomb distortion, the leading term in (\ref{2.23}), $\frac{d\sigma^{B1}}{d\Omega_f}$,  is conventionally replaced by the exact result,
\begin{equation}\label{2.24}
\frac{d\sigma^{\rm coul}}{d\Omega_f}\;=\;\frac{|\bfk_f|}{|\bfk_i|}\;\frac{1}{f_{\rm rec}}\;\frac12\sum_{\sigma_i \sigma_f} |f_{\rm coul}(\sigma_i \sigma_f)|^2,
\end{equation}
where $f_{\rm coul}$ is the scattering amplitude obtained from the phase-shift analysis.
In terms of the spin-conserving (A) and spin-flip (B) amplitudes \cite{Lan},
the contributions from the electronic states with positive (+) or negative (-) helicity are given by
$$f_{\rm coul}(+\,+)\;=\;f_{\rm coul}(-\,-)\;=\;A,$$
\begin{equation}\label{2.25}
f_{\rm coul}(+\,-)\;=\;-\;f_{\rm coul}(-\,+)\;=\;i\;B,
\end{equation}
and further
\begin{equation}\label{2.26}
\frac12 \sum_{\sigma_i \sigma_f} |f_{\rm coul}(\sigma_i \sigma_f)|^2\;=\;|A|^2\;+\;|B|^2.
\end{equation}
This leads to the Born-type cross section formula,
$$\frac{d\sigma^{B1-{\rm type}}}{d\Omega_f}\;=\;\frac{|\bfk_f|}{|\bfk_i|}\;\frac{1}{f_{\rm rec}}  \;\frac12 \sum_{\sigma_i \sigma_f} \left[ \,|f_{\rm coul}|^2\right. $$
\begin{equation}\label{2.27}
\left. +\;2\mbox{ Re}\left\{A_{fi}^{\ast B1} \left(A_{fi}^{\rm vac} + A_{fi}^{\rm vs} + A_{fi}^{\rm box}\right)\right\}\,+\;\frac{d\sigma^{\rm soft}}{d\Omega_f}\right].
\end{equation}
The term proportional to $A_{fi}^{B2}$ does no longer occur, since it is already incorporated into the exact leading term. 

At higher collision energies or larger scattering angles, when the electron-nucleus distance (given approximately by the inverse momentum transfer)
becomes comparable to the nuclear radius, there occur notable differences between $f_{\rm coul}$ and $A_{fi}^{B1}$, even for light nuclei such as $^{12}$C.
This is shown in Fig.1 where $\frac{d\sigma^{\rm coul}}{d\Omega_f}$ is compared to $\frac{d\sigma^{B1}}{d\Omega_f}$ for a collision energy of 240.2 MeV.
It is seen that diffraction affects the PWBA at higher momentum transfer than the phase-shift theory, because 
in the latter the electron is allowed to accelerate in the attractive field of the nucleus, diminishing the distance to the target.

%Fig.1\\
\begin{figure}
\vspace{-1.5cm}
\includegraphics[width=11cm]{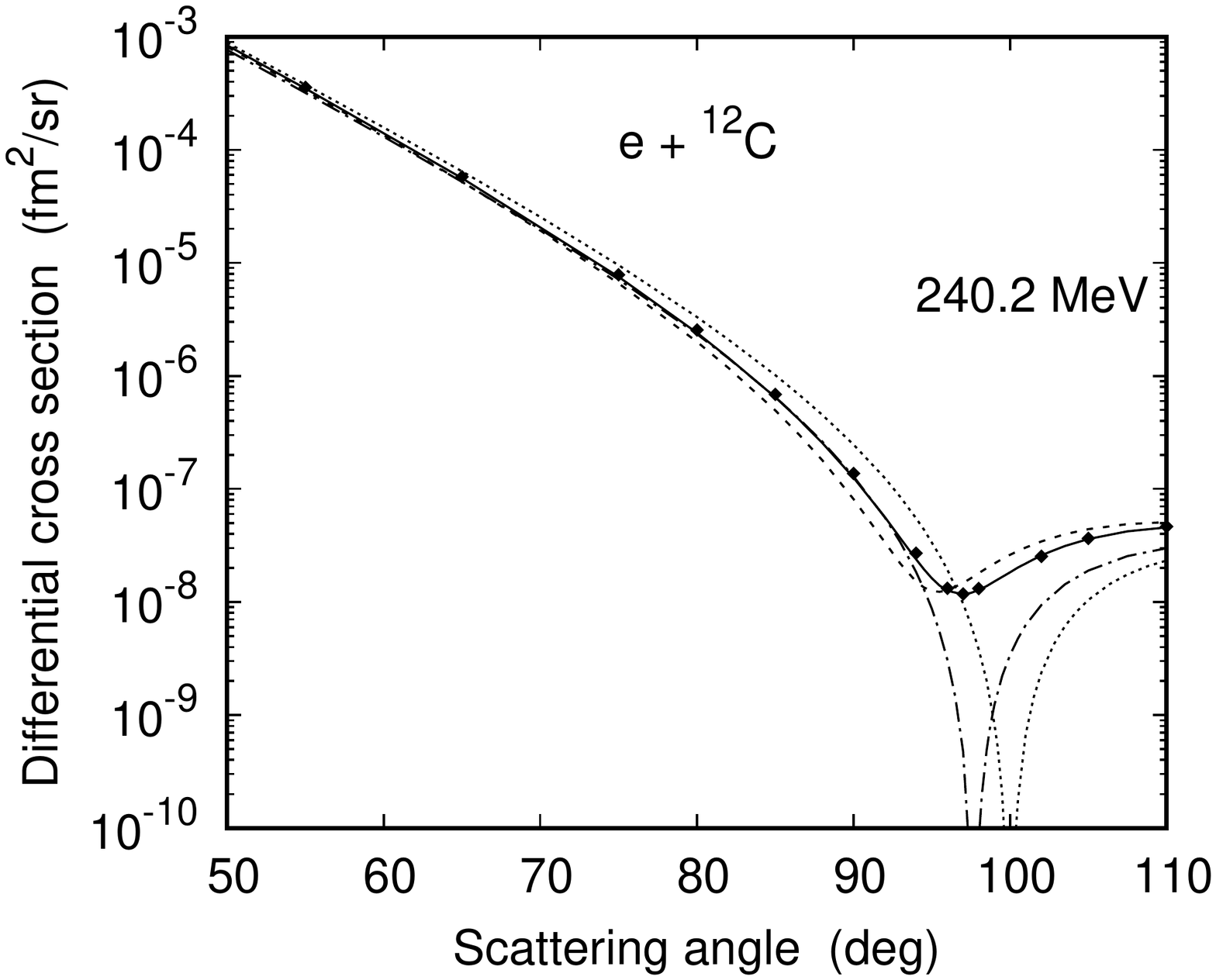}
%\vspace{-0.5cm}
\caption
{
Differential cross section for the elastic scattering of 240.2 MeV electrons from $^{12}$C as a function of scattering angle $\vartheta_f$.
Shown is the PWBA result $d\sigma^{B1}/d\Omega_f$ without $(\cdots\cdots)$ and with $(- \cdot - \cdot -)$ consideration of $q_{\rm eff}$,
as well as the phase-shift results $d\sigma^{\rm coul}/d\Omega_f$ from (\ref{2.24}) with collision energy $E_{i,{\rm kin}}\;\,(- - - -)$ and with $\bar{E}$ instead (----------).
The experimental data ($\blacklozenge$) are from Reuter et al \cite{Reu82}.
}
\end{figure}

In order to improve on the Born amplitude, distortion is conventionally accounted for by shifting the momentum $|\bfq|$, occurring in (\ref{2.20}), to a slightly higher value $q_{\rm eff}$, given by \cite{Reu82}
\begin{equation}\label{2.28}
q_{\rm eff}\;=\;|\bfq|\;\left( 1\,+\;\frac{3Z}{2c}\;\frac{\hbar c}{\sqrt{\frac{5}{3}\,\langle r^2\rangle}\,E_{i,{\rm kin}}}\right),
\end{equation}
with $\hbar c=197.4$ MeV$\,$fm and the kinetic energy $E_{i,{\rm kin}}=E_i-c^2$ in MeV.
The quantity $\sqrt{\langle r^2 \rangle}$ is the root-mean-square charge radius of the nucleus, which for $^{12}$C is  2.47 fm \cite{Reu82,Off91}.
It is seen in Fig.1 that with this prescription, the minimum of the Born theory basically coincides with the experimental position.

The deviation of $\frac{d\sigma^{\rm coul}}{d\Omega_f}$ from experiment \cite{Reu82} near the position of the minimum, seen in Fig.1, can be ascribed to the omission of recoil in the phase-shift analysis.
This contrasts the PWBA result where the different energies of initial and final electronic states can easily be taken into account.

Several ways to include recoil in the phase-shift analysis are considered in the literature. A modification of the Dirac equation is suggested in \cite{Fo59}, and a scaling of the nuclear potential is derived in \cite{CJ88}.
Here we adopt the prescription from \cite{MG64,Ma69} which consists in replacing the kinetic energy $E_{i,{\rm kin}}$
 by an average collision energy $\bar{E}=\sqrt{E_{i,{\rm kin}}E_{f,{\rm kin}}}$ (where $E_{f,{\rm kin}}=E_f-c^2$) when performing the phase-shift analysis.
Modifying  $f_{\rm coul}$ in this way leads to a good agreement with the data concerning the position of the  diffraction minimum.
By inspection of Fig.1 one can also see that at angles $60^\circ \lesssim \theta \lesssim 90^\circ$ this result is very close to the $q_{\rm eff}$-modified PWBA result.
Note that at very small angles (respectively at very low momentum transfer) the $q_{\rm eff}$-prescription fails, and the recoil-modified $f_{\rm coul}$ is closer to the unshifted $A_{fi}^{B1}$ from (\ref{2.20}).

However, even with the above modifications, the mismatch in intensity between the Born theory and the phase-shift analysis persists near and beyond the minimum.
As a consequence, the lowest-order QED correction terms, which in fact are proportional to $|A_{fi}^{B1}|^2$, will be seriously in error in the vicinity of the diffraction minimum where they are most likely to be observed.
Following the suggestion of Maximon \cite{Ma69} to consider Coulomb distortion also in the next-to-leading-order terms, we replace $A_{fi}^{B1}$ with $f_{\rm coul}$ throughout.
This implies that (\ref{2.27}) is exchanged for
$$\frac{d\sigma^{\rm tot}}{d\Omega_f}\;=\;\frac{|\bfk_f|}{|\bfk_i|}\;\frac{1}{f_{\rm rec}}  \;\frac12 \sum_{\sigma_i \sigma_f} \left[ \,|f_{\rm coul}|^2\right.$$
\begin{equation}\label{2.29}
\left. +\;2\mbox{ Re}\left\{ f_{\rm coul}^\ast \left( \tilde{A}_{fi}^{\rm vac}+\tilde{A}_{fi}^{\rm vs}+A_{fi}^{\rm box}\right)\right\}\;+\;\frac{d\tilde{\sigma}^{\rm soft}}{d\Omega_f}\right]
\end{equation}
where the tilde indicates that in (\ref{2.1}), (\ref{2.4}) and (\ref{2.9}), $A_{fi}^{B1}$ is replaced by $f_{\rm coul}$.
This does not affect the cancellation of the IR terms in the vsb contribution.
We note that the replacement implied in (\ref{2.29}) should not be done for perpendicular  spin asymmetry considerations, since this would suppress its changes by the vacuum polarization and by the vsb correction for which $|f_{\rm coul}|^2$ is a common factor.

\section{Results}
\setcounter{equation}{0}

In this section we provide datails of our numerical computations and define the relative changes of the differential cross section due to the QED and the dispersion effects.
Both angular and energy distributions are considered in comparison with experiment.
If not stated otherwise, the reduced energy $\bar{E}$ will be used throughout in $f_{\rm coul}$, and the effective momentum $q_{\rm eff}$ in $A_{fi}^{B1}.$
Finally we discuss the findings of the Jefferson Lab experiment for 362 MeV impact energy
at an observation angle of $61^\circ$ \cite{Jef20},
which is in the region of  the first diffraction minimum.

\subsection{Numerical details}

For obtaining the phase-shift result, the  electronic scattering  state $\psi_i$ is decomposed into partial waves, and for each partial wave the radial Dirac equations
are solved by means of the Fortran code RADIAL from Salvat et al \cite{Sal}. The weighted summation of the corresponding phase shifts
is carried out with the help of  a threefold convergence acceleration \cite{YRW}.
The target potential $V_T$ is generated from the spherical nuclear charge distribution $\varrho_N$ which for $^{12}$C is available in terms of a Fourier-Bessel expansion,
\begin{equation}\label{3.1}
\varrho_N(r)\;=\;\left\{ \begin{array}{ll}
{\displaystyle \sum_{k=1}^N a_k\;j_0\left( \frac{k \pi r}{R_0}\right)}, & r \leq R_0\\
&\\
0,& r>R_0,
\end{array}\right.
\end{equation}
where $j_0$ is a spherical Bessel function, and
the parameters $a_k,\;R_0$ and $N$ are tabulated for $^{12}$C in \cite{VJ}. When 
 explicitly stated, the tabulation from Offerman et al \cite{Off91} is used instead.
This Fourier-Bessel expansion allows for an analytical formula for the target potential,
\begin{equation}\label{3.2}
V_T(r)\;=\;\left\{ \begin{array}{ll}
{\displaystyle -4\pi \sum_{k=1}^N \frac{a_k}{\alpha_k^2}\left[ \frac{\sin (\alpha_kr)}{\alpha_k}\;-\;(-1)^k\right]},& r\leq R_0\\
&\\
{\displaystyle -\;\frac{Z}{r}},& r>R_0,
\end{array}\right.
\end{equation}
where $\alpha_k=\frac{k\pi}{R_0}$.
From the representation (\ref{2.5}), the Coulombic form factor $F_L$ is obtained by means of
\begin{equation}\label{3.3}
F_L(q)=\frac{2 \pi}{Z\,q}\sum_{k=1}^N \frac{a_k}{\alpha_k}\left[ \frac{\sin[(q\!-\!\alpha_k)R_0]}{q-\alpha_k} 
 -\frac{\sin[(q\!+\!\alpha_k)R_0]}{q+\alpha_k} \right],
\end{equation}
where the argument of $F_L$ is identified with $|\bfq|$.
The form factor $F_L$ entering into the correlation function $C^{00}$ for the box diagram  is, however, calculated from the simple formula provided in \cite{FR74} 
(unless stated otherwise), in order to be consistent with the oscillator model for $C^{00}$.
The evaluation of the integral in (\ref{2.17}) is described in \cite{FR74}; however, we have not introduced any further approximations.
In particular, the two values of the integration variable $|\bfq_1|$ where the energy denominator may become zero,
are determined with Newton's method. One of them is close to $\bar{\omega}/c$, the other is of the order of $k_i$.
Near these singularities, the integration over the polar angle $\vartheta_{q_1}$ has to be carried out analytically, while the radial and the azimuthal-angle integrals can be performed numerically. More details are given in \cite{Jaku21}.

The free 4-spinors $u_{k_n}^{(\sigma_n)}$, $n=i,f$, entering into $A_{fi}^{B1},\;A_{fi}^{\rm box}$ and $A_{fi}^{\rm vs}$, pertain to the initial and final helicity eigenstates and are given by
\begin{equation}\label{3.4}
u_{k_n}^{(\pm)}\;=\;\sqrt{\frac{E_n+c^2}{2E_n}}\;{1 \choose c \bfsigma \bfk_n/(E_n+c^2) }\,\chi_{\pm \frac12}
\end{equation}
with $\chi_{\frac12}={1 \choose 0}$ and $\chi_{-\frac12}={0 \choose 1}$. The vector $\bfsigma$ consists of the Pauli matrices. 

The final momentum $k_f$ is determined from the four-momentum conservation,
\begin{equation}\label{3.5}
q\;=\;P_f\,-\,P_i\;=\;k_i\,-\,k_f.
\end{equation}
Upon squaring this equation and using Eq.(10.43) and Eq.(10.48) of \cite{Gro}, one obtains, defining the scattering angle $\vartheta_f = \angle (\bfk_i,\bfk_f),$
$$|\bfk_f|\;=\;\frac{1}{a}\;\left( b\,+\,\sqrt{b^2-ad}\right),$$
$$a\;=\;(Mc\,+\,E_i/c)^2 \;-\;(|k_i|\cos \vartheta_f)^2,$$
$$ b\;=\;(Mc \cdot E_i/c \,+\,c^2)\,|k_i|\cos\vartheta_f,$$
\begin{equation}\label{3.6}
d\;=\;-\left( (Mc)^2\,-\,c^2\right)\;\left( (E_i/c)^2 \,-\,c^2\right),
\end{equation}
where $M=12\times 1836$ in atomic units, and we have assumed that initially the nucleus is at rest.
In all calculations, the $z$-axis is determined by $\bfk_i$ and the scattering plane coincides with the $(x,z)$-plane, such that $\bfk_f=|\bfk_f|(\sin \vartheta_f,0,\cos \vartheta_f).$

%Fig.2
\begin{figure*}[t]
\centering
\begin{tabular}{cc}
\hspace{-1cm}\includegraphics[width=.7\textwidth]{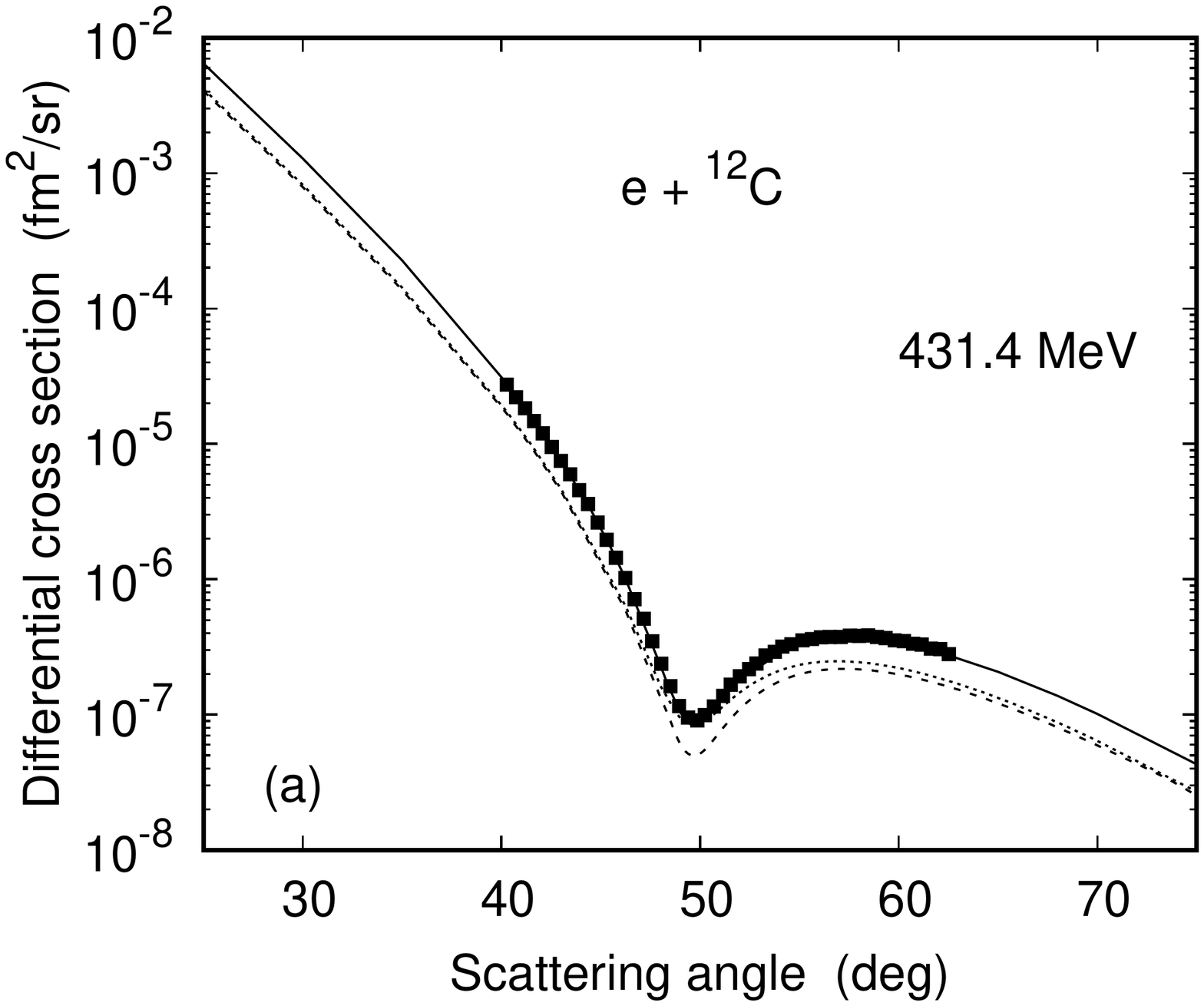}&
\hspace{-3.0cm} \includegraphics[width=.7\textwidth]{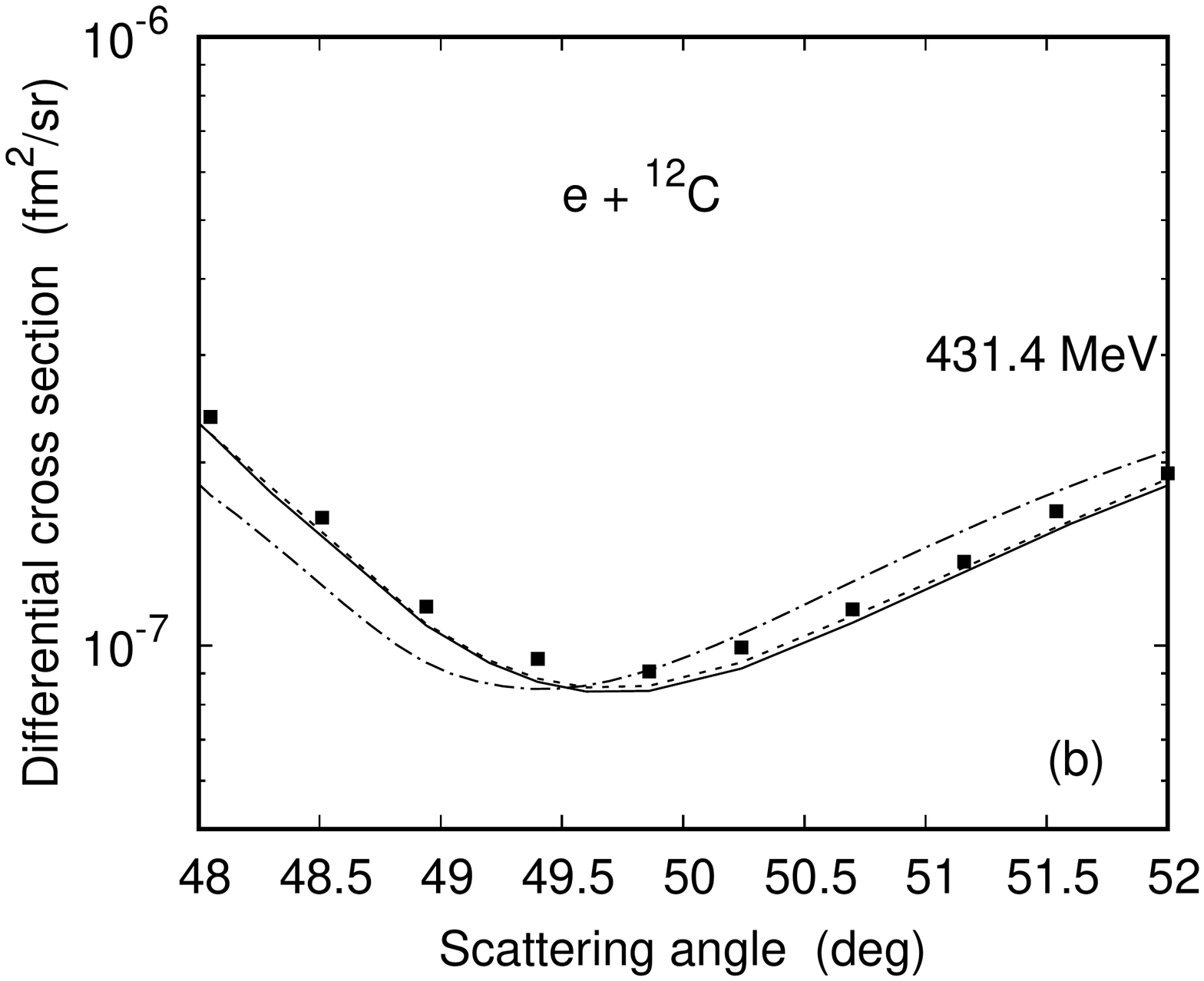}\\
\hspace{-1cm}\includegraphics[width=.7\textwidth]{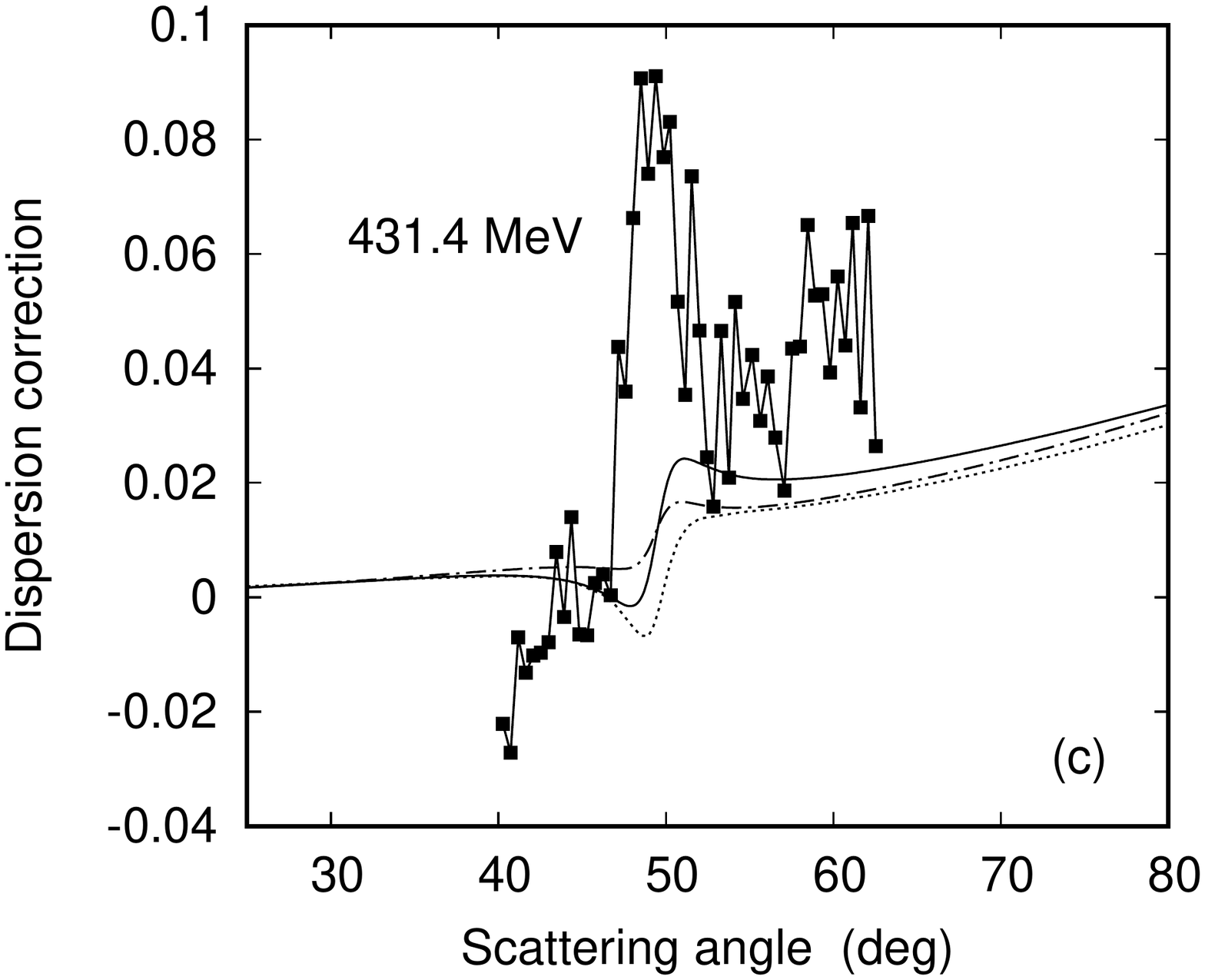}&
\hspace{-3.0cm} \includegraphics[width=.7\textwidth]{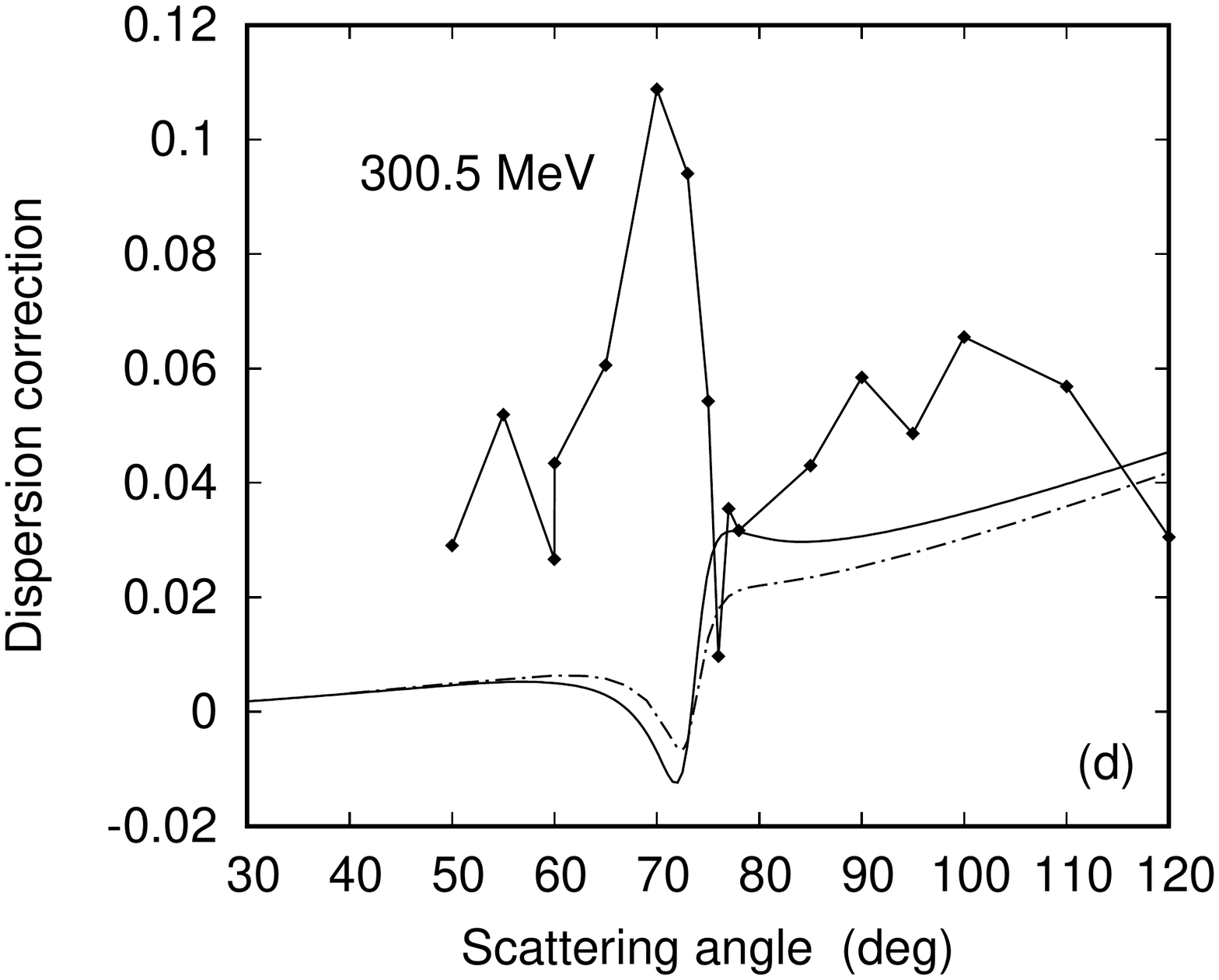}
\end{tabular}
\caption{
Differential cross section from 431.4 MeV $e+^{12}$C collisions and the cross section change by dispersion for 431.4 MeV and 300.5 MeV as a function of scattering angle $\vartheta_f$.
Shown in (a) is $d\sigma^{\rm coul}/d\Omega_f$ (----------) and $d\sigma_{\rm tot}/d\Omega_f$ ($- - - -)$.
Included are results from the Born-type theory (\ref{2.27}) ($\cdots\cdots$). The experimental data $(\blacksquare$) are from Offermann et al \cite{Off91}. $\Delta E/E =2 \times 10^{-4}$.
Shown in (b) is the enlarged region of the minimum. Included are results from the phase-shift theory using the energy $E_{i,{\rm kin}} \;\,(-\cdot - \cdot -)$, and when the dispersion correction is included in the cross section according to (\ref{2.29})($----$).
Shown in (c) is $\Delta \sigma$ from the dispersion effect for 431.4 MeV: $\Delta \sigma_{\rm box}$ (---------),
when in addition $F_L$ in $C^{00}$ is treated exactly ($-\cdot -\cdot -$), and the Born-type theory
$(\cdots\cdots)$. The experimental data ($\blacksquare$) are from Offermann et al \cite{Off91}.
Shown in (d) is $\Delta\sigma$ from the dispersion effect for 300.5 MeV: $\Delta \sigma_{\rm box}$ (--------) and when in addition $F_L$ in $C^{00}$ is treated exactly ($-\cdot-\cdot -)$. The experimental data ($\blacklozenge$) are from Reuter et al \cite{Reu82}.
}
\end{figure*}

\subsection{Angular dependence}

A comparison of the cross section with experiment
from Offermann et al \cite{Off91}
 for 431.4 MeV electron impact  is provided in Fig.2a.
Like in Fig.1, the experimental data 
 are corrected for the QED effects in order to reproduce the
phase-shift results away from the diffraction minimum, if calculated from a suitably fitted nuclear charge distribution.
Again, the recoil-corrected phase-shift analysis agrees very well with the cross section measurements.
Also shown are results where all corrections are included.
When the formula (\ref{2.29}) is used, where Coulomb distortion is fully accounted for, the cross section is lowered by about 40\%.
In contrast, if the Born-type  prescription (\ref{2.27}) is applied,
the corrections reduce to a few percent near the minimum at $50^\circ$ (since $A_{fi}^{B1}=0$ at the minimum position).

Fig.2b shows an enlarged region around $50^\circ$.
In addition to $\frac{d\sigma^{\rm coul}}{d\Omega_f}$, displayed in Fig.2a, we have included the phase-shift result without consideration of recoil, which clearly shows the mismatch with the data.
Also provided is the modified cross section resulting from adding  the dispersion correction  to the phase-shift result.
Clearly, its effect is too small to compensate the underprediction of the data.
In the minimum at $\vartheta_f=49.86^\circ$ the deviation of the dispersion-corrected theory from the data is 5.6\%.

In order to quantify the QED and dispersion corrections
 it is of advantage to define  the relative change of the cross section 
 according to \cite{Tsa61,MT00}
\begin{equation}\label{3.7}
\Delta \sigma\;=\;\frac{d\sigma/d\Omega_f}{d\sigma^{\rm LO}/d\Omega_f}\;-\;1,
\end{equation}
where $\frac{d\sigma}{d\Omega_f}$ denotes the cross section in which any QED effect or dispersion  is included, while the denominator represents  the leading-order (LO) cross section. 
Unless stated otherwise, $d\sigma^{\rm LO}/d\Omega_f$ will always be identified with $d\sigma^{\rm coul}/d\Omega_f$ from (\ref{2.24}).

The dispersion correction
 $\Delta \sigma_{\rm box}$ is calculated from the Coulomb-distortion theory, (\ref{3.7}) with the use of (\ref{2.29}), by ignoring the contributions from $\tilde{A}_{fi}^{\rm vac},\;\tilde{A}_{fi}^{\rm vs}$ and from the soft bremsstrahlung.
In Fig.2c,
 $\Delta \sigma_{\rm box}$ is compared to the experimental change $\Delta \sigma_{\rm exp}$ for which $d\sigma/d\Omega_f$ in the numerator of (\ref{3.7}) refers to the QED-corrected data points.
As anticipated from Fig.2b, $\Delta \sigma_{\rm exp}$ increases to 8\% in the vicinity of the diffraction minimum, set against the experimental accuracy of about 2\%.

In order to display the influence of different models we have included two more estimates for the dispersion correction: (i) the Born-type theory resulting from (\ref{2.27}) and (ii) the Coulomb-distortion theory where in the correlation function $C^{00}$ the exact form factor $F_L$ according to (\ref{3.3}) is employed.
The latter model may be viewed as an accuracy test of the Friar and Rosen harmonic oscillator model for $C^{00}$.
In their work \cite{FR74} they have tested their model in a different way by
improving on the closure approximation (for an energy of 375 MeV) with the help of an expansion in terms of $\omega_n-\bar{\omega}$. They have  found (as we do for model (ii)) that the negative excursion of $\Delta \sigma_{\rm box}$ has disappeared at the expense of reducing the positive excursion near the minimum.

It is obvious from Fig.2c that below the onset of diffraction, all models lead to the same result. Strong deviations occur in the vicinity of the minimum and persist at still larger angles.
However, none of these model-dependent changes can explain the large values of $\Delta \sigma_{\rm exp}$ in the diffraction region.

Fig.2d shows the dispersion effect for an energy of 300.5 MeV in comparison with the experiments from Reuter et al \cite{Reu82}. For this energy, the deviation of 
the various models for dispersion from the QED-corrected data
is even higher, above 10 percent in the maximum.
The same situation persists for even lower energies. 
We therefore ascribe the deviations between $\Delta \sigma_{\rm exp}$ and $\Delta \sigma_{\rm box}$ partly to an incorrect account of the QED effects when extracting the plotted experimental cross sections from the measured raw data. This conjecture is substantiated below.

%Fig.3
\begin{figure*}[t]
\centering
\begin{tabular}{cc}
\hspace{-1cm}\includegraphics[width=.7\textwidth]{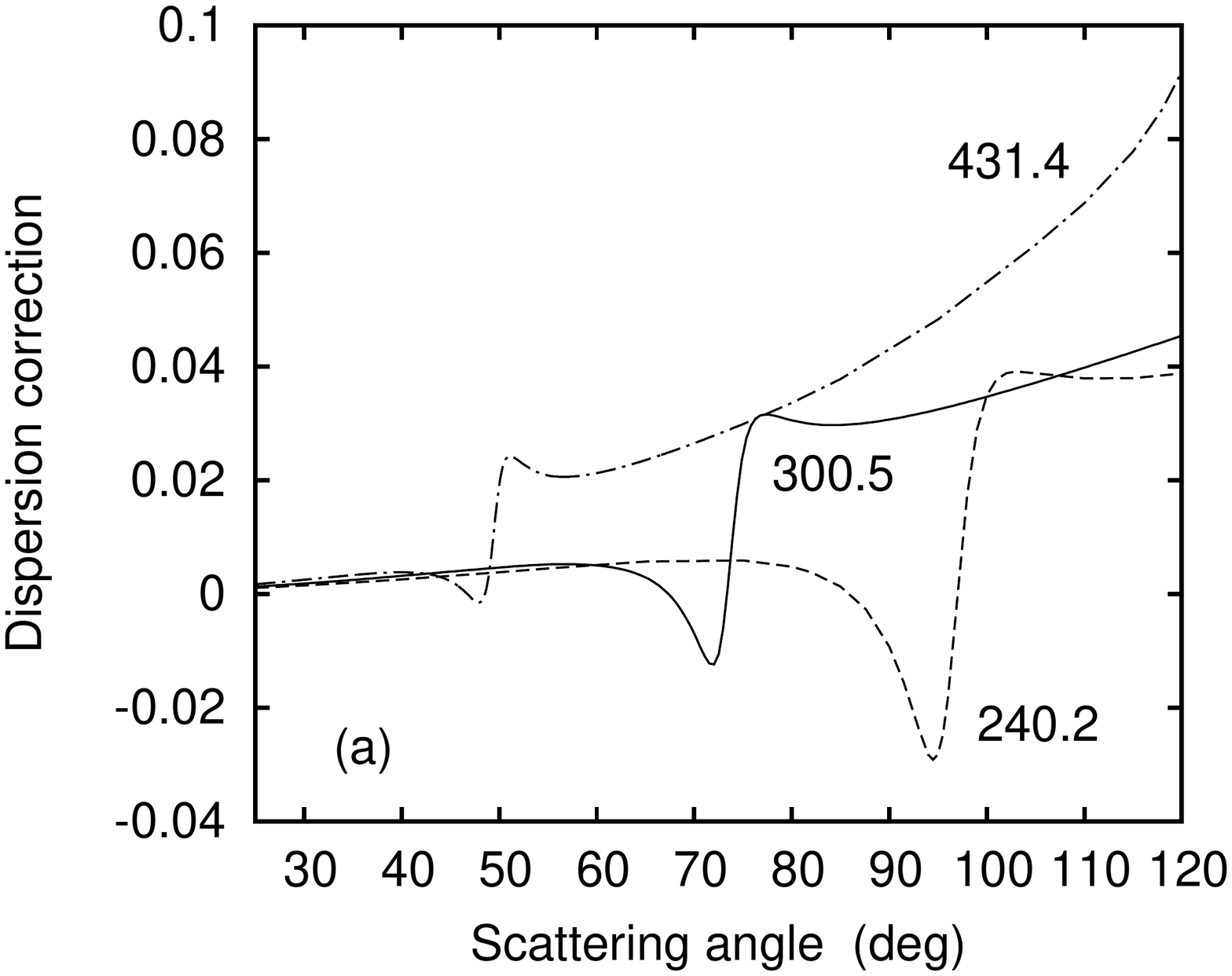}&
\hspace{-3.0cm} \includegraphics[width=.7\textwidth]{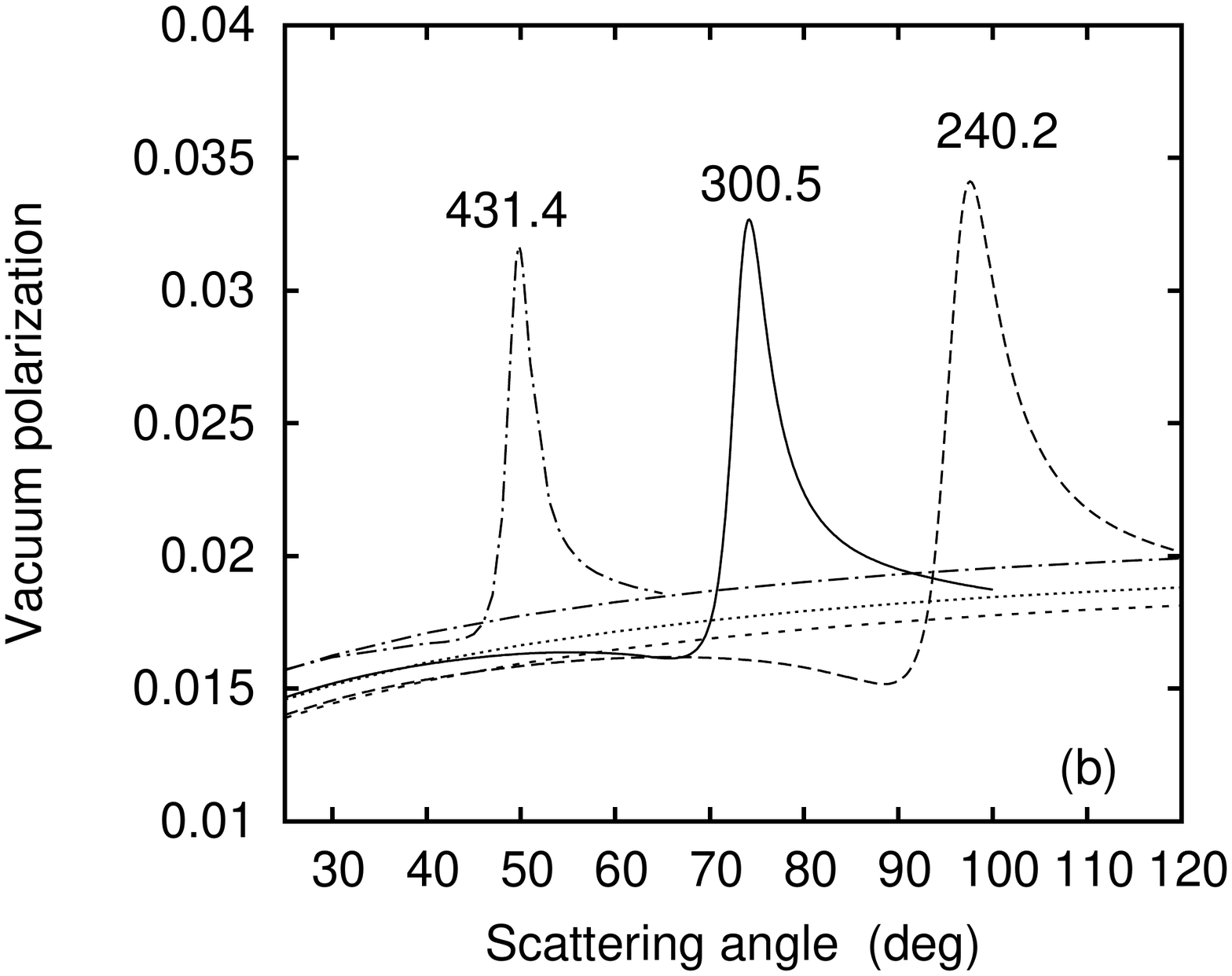}\\
\hspace{-1cm}\includegraphics[width=.7\textwidth]{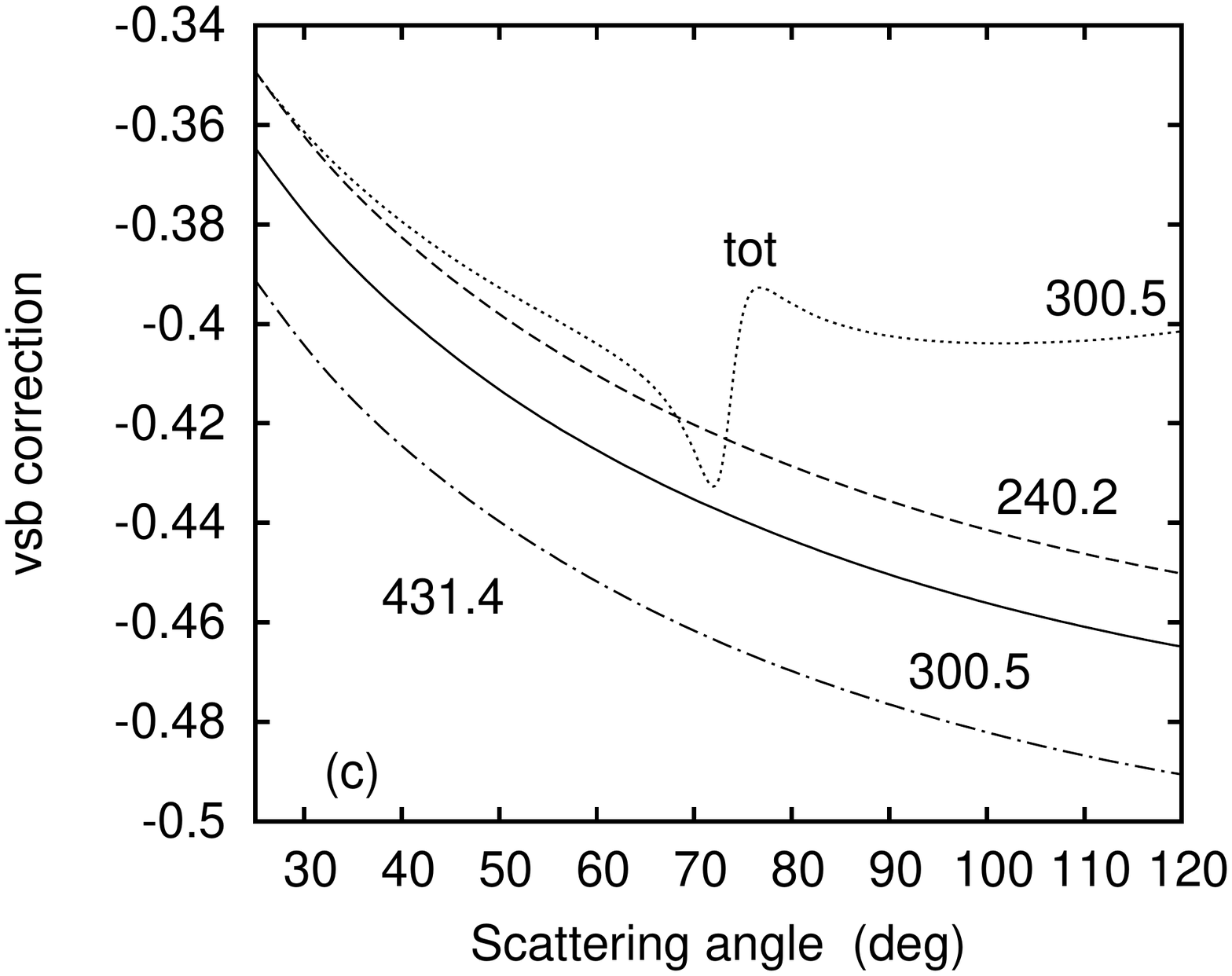}&
\hspace{-3.0cm} \includegraphics[width=.7\textwidth]{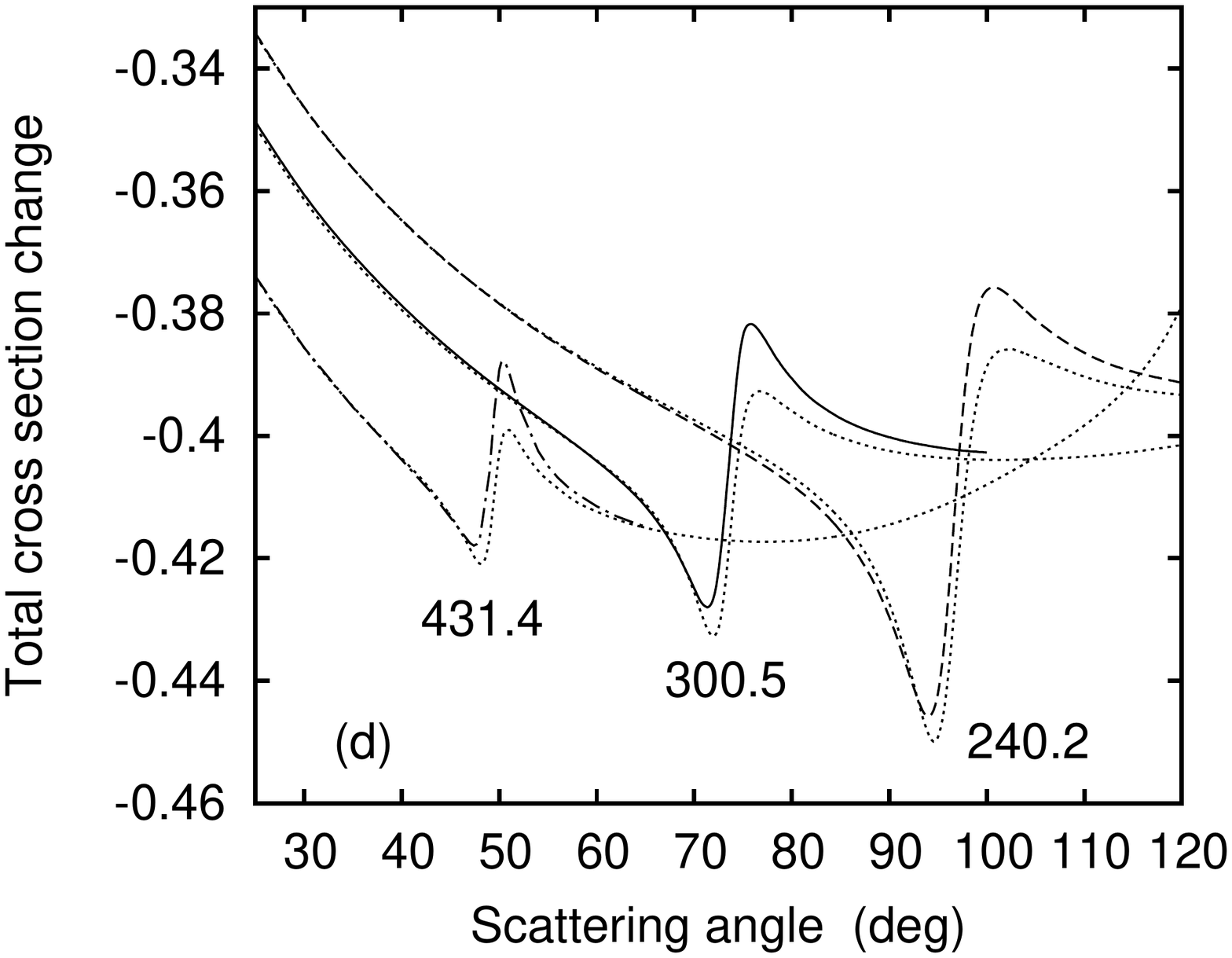}
\end{tabular}
\caption{
Change $\Delta \sigma$ of the differential cross section from elastic $e+^{12}$C collisions (a) by dispersion, (b) by vacuum polarization, (c) by the vsb correction and (d) including all effects, as a function of scattering angle $\vartheta_f$.
In (a) and (c), at the collision energies  240.2 MeV ($----$), 300.5 MeV (---------) and 431.4 MeV $(-\cdot -\cdot -$),
 these results are derived from the Coulomb distortion theory (\ref{2.29}).
Shown in (b) is $\Delta \sigma_{\rm vac}$ for 240.2 MeV ($-  -- -$), 300.5 MeV (--------) and 431.4 MeV ($-\cdot - \cdot -$, upper line). Included is $\Delta \sigma_{\rm vac}^{B1}$ for 240.2 MeV ($----$), 300.5 MeV $(\cdots\cdots$) and 431.4 MeV ($-\cdot - \cdot -$, lower line).
In (d), the total change $\Delta \sigma_{\rm ex}$ from (\ref{3.9}) is shown for 240.2 MeV $(----$), 300.5 MeV (--------) and 431.4 MeV $(-\cdot -\cdot -$). The results for $\Delta \sigma_{\rm tot}$ from (\ref{2.29}) are also shown ($\cdots\cdots)$. This latter result for 300.5 MeV is included in (c)
 ($\cdots\cdots)$.
}
\end{figure*}

Fig.3a provides the angular dependence of the dispersion correction, calculated with the Coulomb-distortion theory (\ref{2.29}), for three collision energies, 240.2 MeV,  300.5 MeV and 431.4 MeV, for which  experimental data are existing.
Clearly, the diffraction structures shift to smaller angles at higher energies, in accordance with a fixed momentum transfer in the form factor which determines the cross section minimum.
However, $|\Delta \sigma_{\rm box}|$ decreases strongly with energy in the region of the cross section minimum. At still larger angles the dispersion effect is most pronounced for the highest $E_i$ as expected, increasing up to 10\% for 431.4 MeV at $120^\circ$.

The difference between the Born approximation and an exact treatment can easily be shown in the case of vacuum polarization.
In Fig.3b we compare the cross section change (\ref{3.7}) as obtained from the exact theory according to
\begin{equation}\label{3.8}
\frac{d\sigma^{\rm vac}}{d\Omega_f}\;=\;\frac{|\bfk_f|}{|\bfk_i|}\;\frac{1}{f_{\rm rec}}\;\frac12\sum_{\sigma_i \sigma_f} |f_{\rm vac}|^2,
\end{equation}
where $f_{\rm vac}$ is the scattering amplitude obtained from the phase-shift analysis with central potential $V_T + U_e$,
with the Coulomb-distorted PWBA result by using (\ref{2.29}) and retaining in the next-to-leading-order terms only the contribution from $\tilde{A}_{fi}^{\rm vac}$.
Due to the proportionality of the differential cross section to $|f_{\rm coul}|^2$ in the latter case, 
any diffraction effect cancels out in (\ref{3.7}), and the resulting change $\Delta \sigma_{\rm vac}^{B1}$ is smoothly increasing with scattering angle and with collision energy.
On the other hand, $\Delta \sigma_{\rm vac}$ obtained from the exact theory shows large excursions in the region of the cross section minima,
up to 3\% independent of energy, their width decreasing with $E_i$.

The change induced by the vsb correction is shown in Fig.3c.
Here, the upper limit $\omega_0$ of the bremsstrahlung radiation is set equal to $\omega_0=E_{i,{\rm kin}}\,\frac{\Delta E}{E}$, where the experimental spectrometer resolution $\frac{\Delta E}{E}=2 \times 10^{-4}$ is taken from Offermann et al \cite{Off91}.
Like for vacuum polarization, the Coulomb distortion theory obtained from (\ref{2.29}) by omitting $\tilde{A}_{fi}^{\rm vac}$ and $A_{fi}^{\rm box}$, leads to a monotonous increase of the magnitude of this QED correction with $\vartheta_f$ and $E_i$.
In concord with the findings for vacuum polarization, it is expected that any exact consideration of the vsb effect will show  structures in the vicinity of the cross section minima.
Therefore, the subtraction of the vsb effect as a smooth background is considered to be erroneous.

The total changes by the QED effects and by dispersion are depicted in Fig.3d for the three collision energies.
Shown are the Coulomb distortion results by either  treating vacuum polarization in Born according to (\ref{2.29}), or by considering it exactly by using instead
$$\frac{d\sigma^{\rm ex}}{d\Omega_f}\;=\;\frac{|\bfk_f|}{|\bfk_i|}\;\frac{1}{f_{\rm rec}}\;\frac12\sum_{\sigma_i \sigma_f} \left[ |f_{\rm vac}|^2 \right.$$
\begin{equation}\label{3.9}
\left. +\;2\,\mbox{Re}\left\{f_{\rm coul}^\ast (\tilde{A}_{fi}^{\rm vs} + A_{fi}^{\rm box})\right\} +\;\frac{d\tilde{\sigma}^{\rm soft}}{d\Omega_f}\right].
\end{equation}
The corrections $\Delta \sigma_{\rm tot}$ relating to (\ref{2.29}), like those from the use of (\ref{3.9}),  increase in magnitude with $\vartheta_f$ for small angles, being dominated by the vsb effect (see Fig.3c where $\Delta \sigma_{\rm tot}$ is included for 300.5 MeV).
The structures in $\Delta \sigma_{\rm tot}$ near the minima, but also the increase at very large angles, relate to the dispersion effect.
The exact consideration of vacuum polarization, $\Delta \sigma_{\rm ex}$, leads to a shift of the structures to smaller angles,
combined with a filling of the minima and an enhancement of the maxima.
At the positions of the cross section minima, the deviations between $\Delta \sigma_{\rm tot}$ and $\Delta \sigma_{\rm ex}$ amount to $3-4\%$.

%Fig.4\\
\begin{figure}
\vspace{-1.5cm}
\includegraphics[width=11cm]{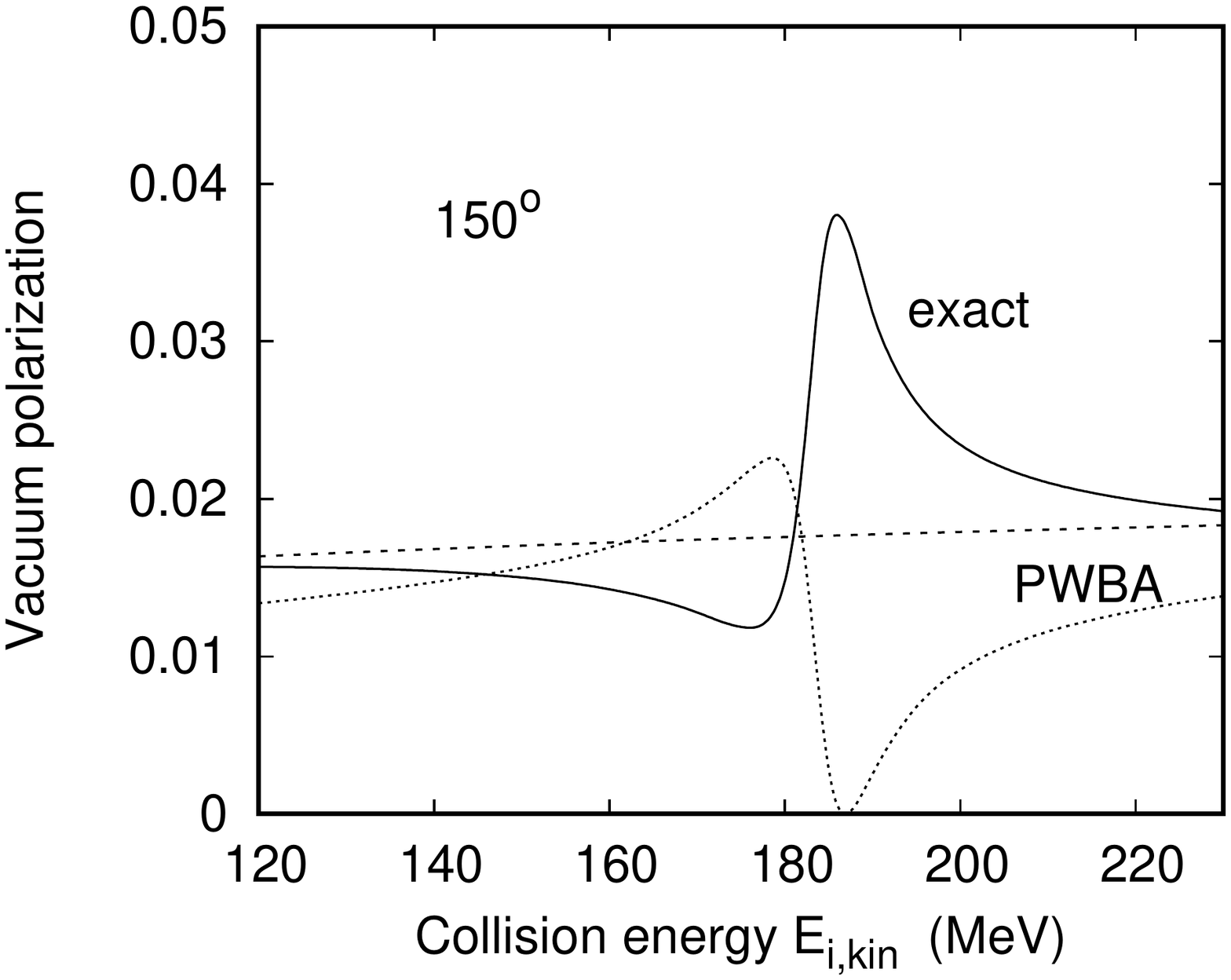}
%\vspace{-0.5cm}
\caption
{
Change $\Delta \sigma$ of the differential cross section from $e+^{12}$C collisions by vacuum polarization at $\vartheta_f=150^\circ$ as a function of collision energy $E_{i,{\rm kin}}$.
Shown is $\Delta\sigma_{\rm vac}$ (--------) and $\Delta \sigma_{\rm vac}^{B1}$ from (\ref{2.29}) $(----)$.
Included are results from the Born-type theory (\ref{2.27})  ($\cdots\cdots)$. 
}
\end{figure}

\subsection{Energy dependence}

In this subsection the Born result from (\ref{2.23})  is included in our calculations of the cross section changes.
Fig.4 displays the energy dependence of the correction  from vacuum polarization at a backward scattering angle, $150^\circ$.
At this angle, the diffraction minimum of the cross section is near 175 MeV.
The exact consideration of vacuum polarization, $\Delta \sigma_{\rm vac}$, shows structures similar to those present in the angular distribution (see Fig.3b).
The results from the Coulomb-distorted Born theory, $\Delta \sigma_{\rm vac}^{B1}$, are monotonously increasing with $E_i$ and can be viewed in terms of a mean cross section change across the diffraction structure.
If the PWBA theory (\ref{2.23}), with $d\sigma^{\rm LO}/d\Omega_f$ in (\ref{3.7}) identified with the Born cross section $\frac{d\sigma^{B1}}{d\Omega_f}$, were used instead, the identical smooth line would be obtained.
This is also true for the vsb correction from Fig.3c, where the PWBA theory reproduces the smooth lines.
If, on the other hand, the Born-type theory (\ref{2.27})  is applied, the structures persist, but they differ notably from those of $\Delta \sigma_{\rm vac}$, most obviously by a change in sign.

In Fig.5 the energy dependence of the cross section with its QED and dispersion corrections is shown for an angle of $90^\circ$.
The experimental data included in Fig.5a result, when not measured at $90^\circ$, from a spline interpolation of adjacent data points.
The recoil-modified phase-shift result explains the measurements well. When all corrections are included, the cross section is lowered by $36-42\%$ if (\ref{3.9}) is used, while
the corrections tend to zero in the vicinity of the minimum if (\ref{2.27})  is applied.

Fig.5b displays the modifications of the cross section by vacuum polarization, treated exactly, and by the dispersion effect $\Delta \sigma_{\rm box}$  according to (\ref{2.29}).
The two effects tend to compensate each other in the region $240-250$ MeV, but lead to a positive change of the cross 
section between 2-6\% at lower or higher energies.

The total change and its dominant contribution, the vsb correction, are displayed in Fig.5c. The structure in $\Delta \sigma_{\rm ex}$ near 250 MeV results basically from dispersion, with some modification from the vacuum polarization.
If, on the other hand, the Born-type theory (\ref{2.27})  is used, not only the total but also the vsb correction shows a diffraction structure, which, however, decreases to zero in its maximum.

%Fig.5
\begin{figure*}[t]
\centering
\begin{tabular}{cc}
\hspace{-1cm}\includegraphics[width=.7\textwidth]{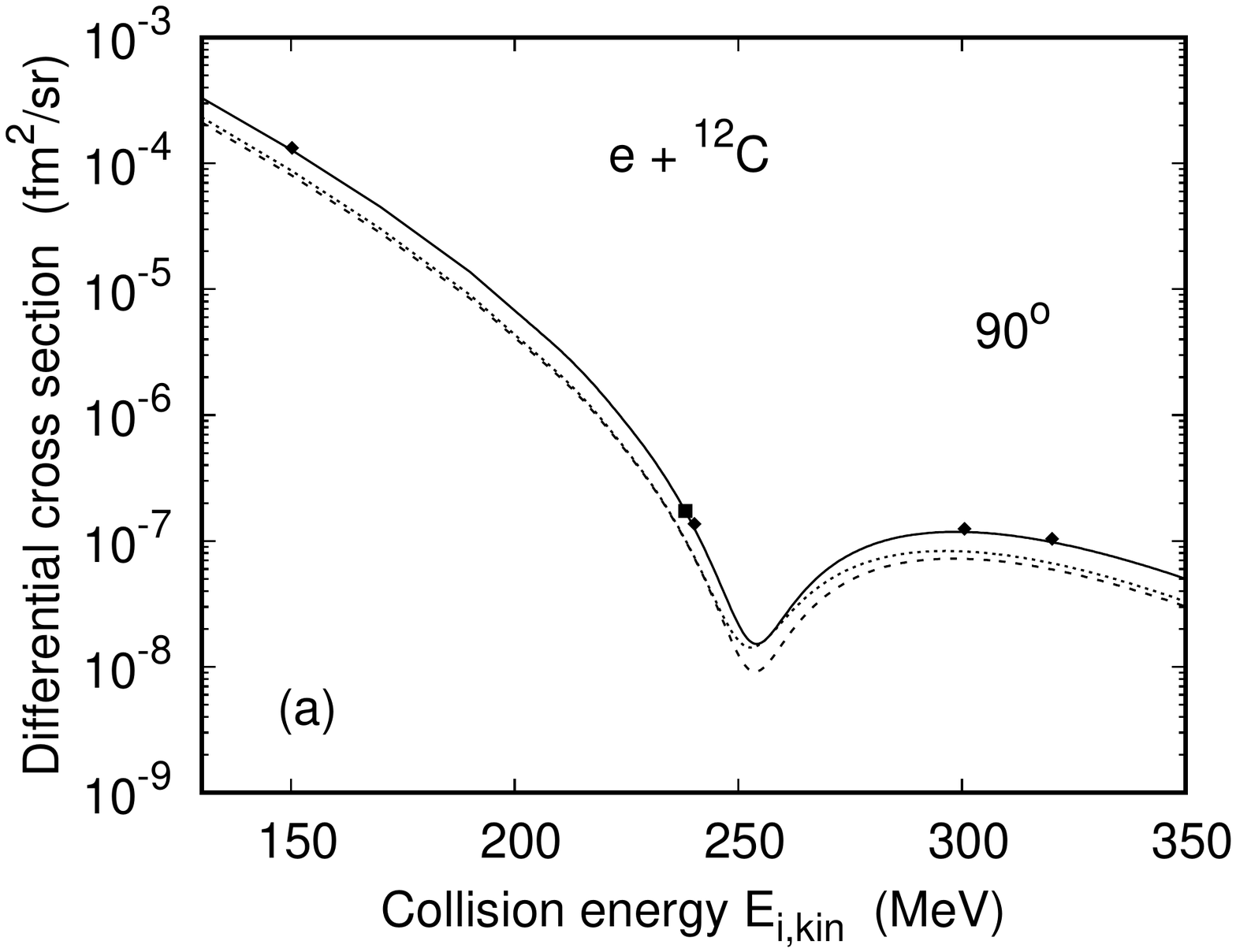}&
\hspace{-3.0cm}\includegraphics[width=.7\textwidth]{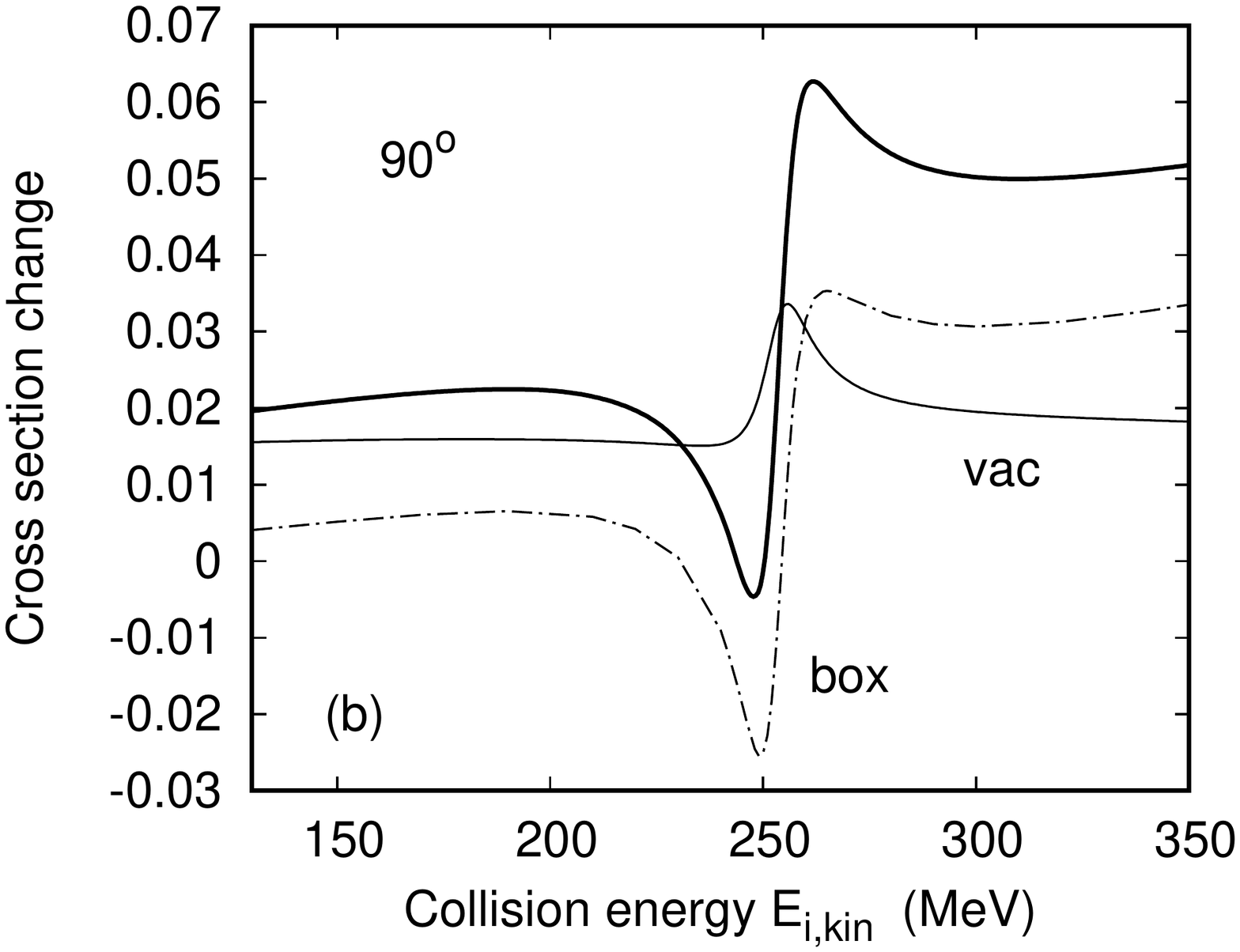}\\
\hspace{-1cm}\includegraphics[width=.7\textwidth]{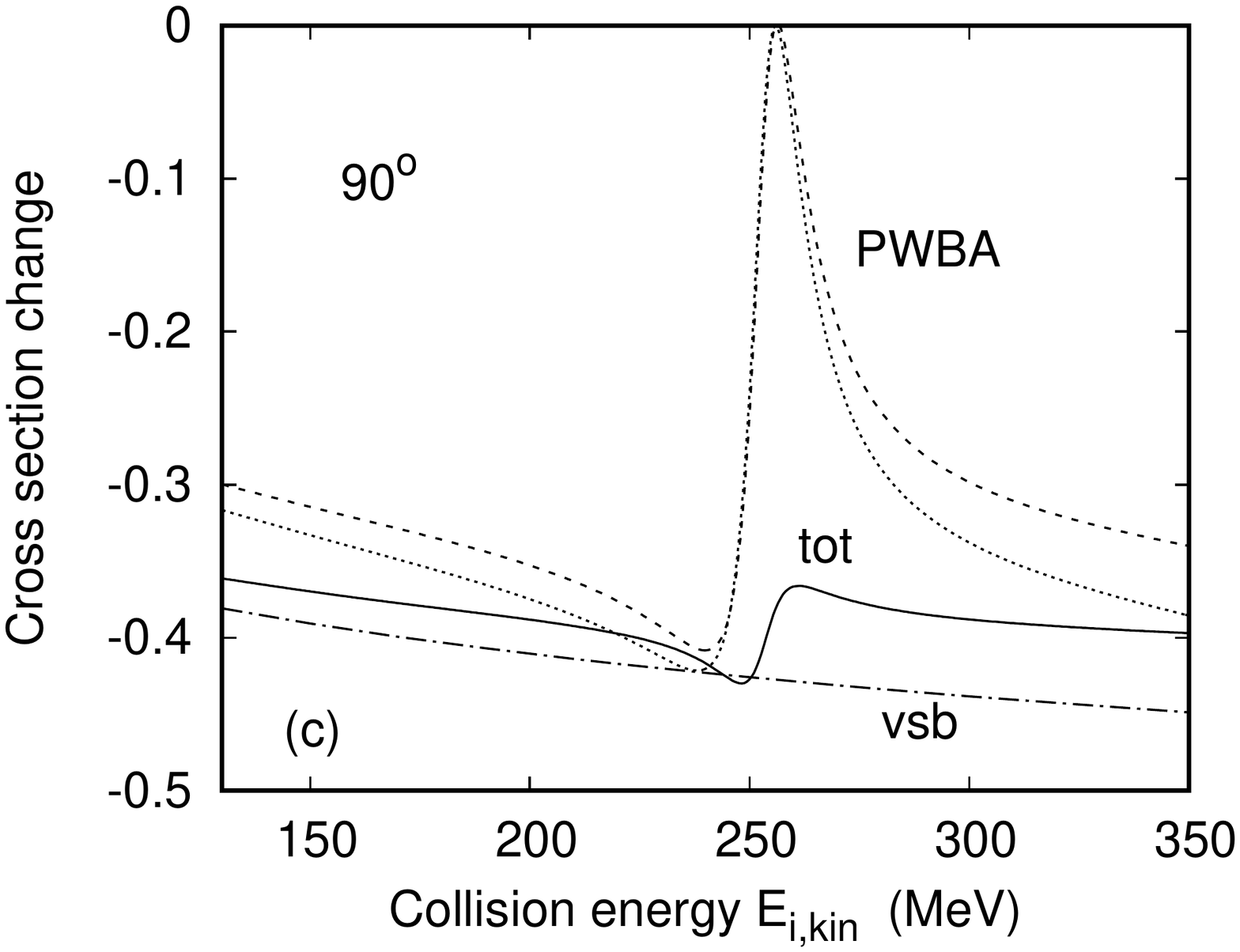}&
\end{tabular}
\caption{
(a) Differential cross section from elastic $e+^{12}$C collisions at $\vartheta_f=90^\circ$ and the cross section change $\Delta \sigma$ (b) from vacuum polarization and the dispersion effect and (c) from the vsb correction and from the combined effecs
as a function of collision energy $E_{i,{\rm kin}}$.
Shown in (a) is $d\sigma^{\rm coul}/d\Omega_f$  (----------) and the total cross section from (\ref{3.9}) ($----$) and from the Born-type theory (\ref{2.27}) ($\cdots\cdots$). The experimental data  are from Reuter et al ($\blacklozenge$ \cite{Reu82}) and from Offermann  et al ($\blacksquare$ \cite{Off91}).
Shown in (b) is $\Delta\sigma_{\rm vac}$ (--------, thin line) and $\Delta \sigma_{\rm box} \;\,(-\cdot -\cdot -$) as well as their sum (--------, thick line).
In (c) the vsb results from (\ref{2.29}) ($-\cdot-\cdot -$) and $\Delta\sigma_{\rm ex}$ (-------------) are compared to the Born-type results  for the vsb effect ($\cdots\cdots$) and for the total correction ($----$).
$\Delta E/E=2.5 \times 10^{-4}$, corresponding to the experimental resolution of \cite{Reu82}.
}
\end{figure*}
\subsection{The Jefferson Lab experiment}

This experiment was carried out at 362 MeV for a single angle. By working in the effective momentum approximation where a modified $q_{\rm eff}$ is defined, which considers simultaneously recoil and distortion \cite{Gue99}, it was found that the cross section minimum is obtained for $61^\circ$,
such that this angle was selected for the measurement \cite{Jef20}.
In contrast to the reduced data tabulated by Reuter et al \cite{Reu82} or Offermann et al \cite{Off91}, the measured cross section is provided without any subtraction of QED corrections. This allows for a straightforward comparison with our QED- and dispersion-corrected theory.

Let us first investigate the precise position of the minimum at 362 MeV with our method of including distortion and recoil. We consider this method superior to the effective momentum approximation, since it is based on the phase-shift analysis and not on the PWBA. Our previous results have shown that the minimum position remains basically unchanged when the QED and dispersion corrections are added to the phase-shift result,
both in the angular as well as in the energy distributions.
Therefore we can  use the recoil-corrected phase-shift theory to determine for each energy $E_{i,{\rm kin}}$
the scattering angle $\vartheta_{\rm min}$ which leads to the smallest cross section, and  subsequently to compare these values with the experimental ones obtained by a spline interpolation of adjacent data points.
This is done in Fig.6a,  and it is seen that our results agree with the  angular position derived from the 
experiments by Offermann et al \cite{Off91} and Reuter et al \cite{Reu82} within 0.2\%. On the other hand, the Jefferson Lab datum point is  well above theory, suggesting that at 362 MeV, the minimum is at $60.0^\circ$ and not at $61^\circ.$
Fig.6b shows the corresponding cross section $\frac{d\sigma^{\rm coul}}{d\Omega_f}(\vartheta_{\rm min})$ as a function of energy.
It is seen that beyond 300 MeV, experiment is considerably underestimated by theory.

%Fig.6\\
\begin{figure}
\vspace{-1.5cm}
\hspace*{-1.0cm}\includegraphics[width=13cm]{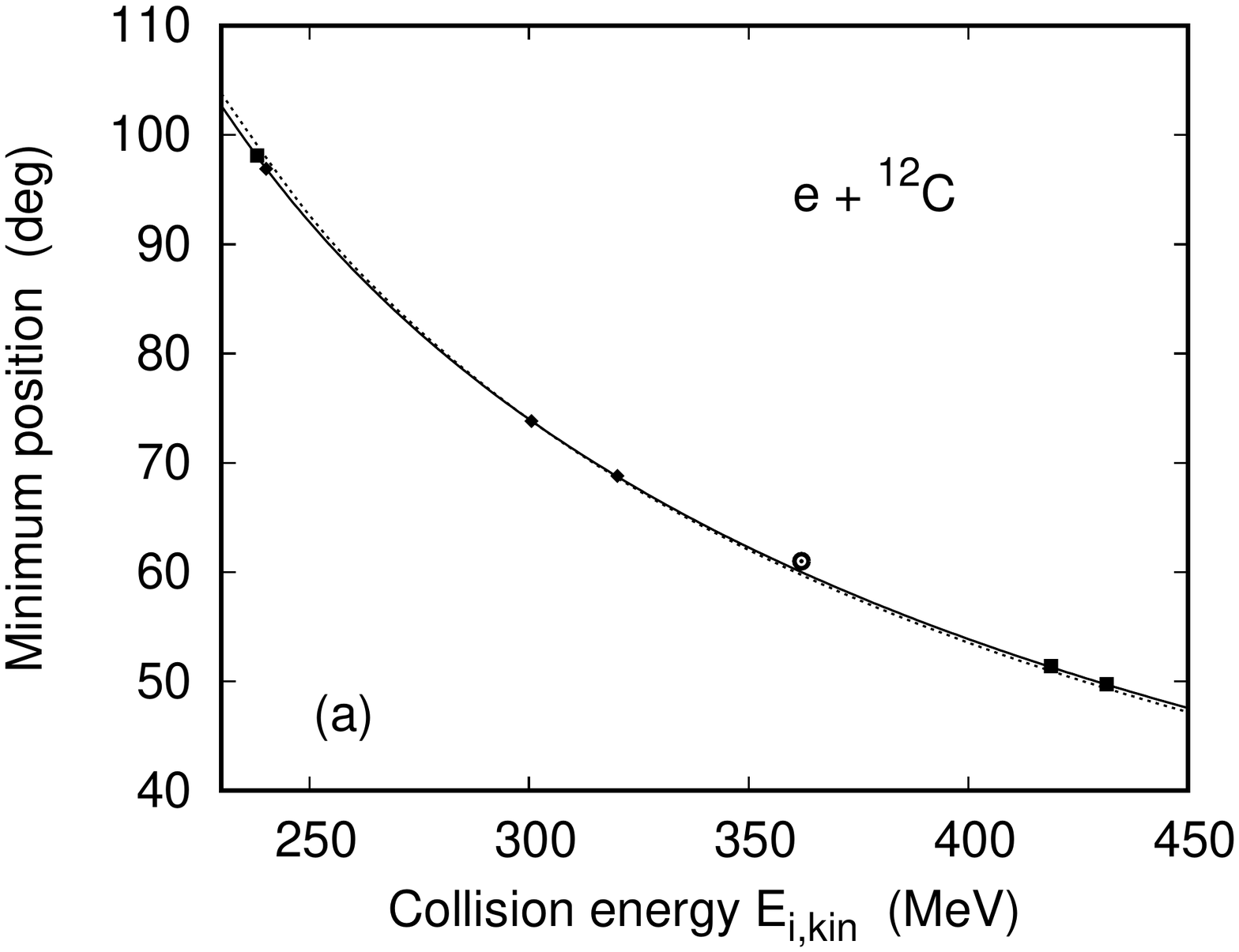}
\vspace{-1.5cm}
\vspace{-0.5cm}
\hspace*{-1.0cm}\includegraphics[width=13cm]{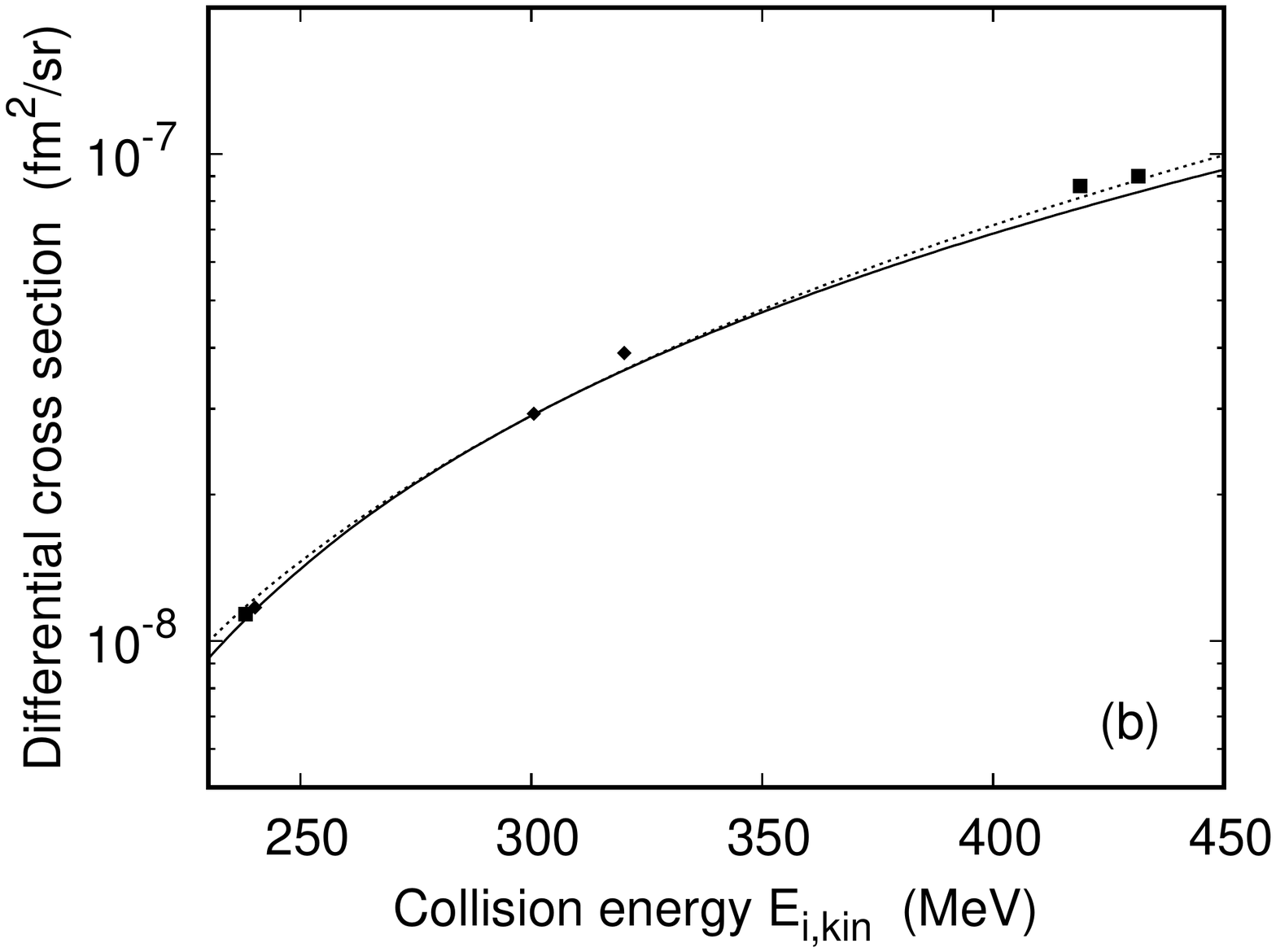}
\caption
{
(a) Position $\vartheta_{\rm min}$ of the first diffraction minimum in elastic $e+^{12}$C collisions and (b) differential cross section in this minimum as a function of collision energy $E_{i,{\rm kin}}$.
Shown in (a) are results from the recoil-corrected phase-shift theory (--------------), and from fixing the four-momentum transfer at 1.8154 fm$^{-1}$ ($\cdots\cdots)$.
Shown in (b) are $\frac{d\sigma^{\rm coul}}{d\Omega_f}(\vartheta_{\rm min})$ (--------) and  $\frac{d\sigma^{\rm coul}}{d \Omega_f}(\vartheta')\;\, (\cdots\cdots$),
where $\vartheta'$ relates to the angle for fixed momentum transfer from (a).
The experimental data are from Reuter et al ($\blacklozenge$ \cite{Reu82}) and from Offermann et al ($\blacksquare$ \cite{Off91}). In (a) the Jefferson Lab datum point is included ($\odot$ \cite{Jef20}). 
}
\end{figure}

It is well understood that diffraction and hence the
minimum position is basically determined by the form factor $F_L$ from (\ref{3.3}). In turn, $F_L$ depends only on the momentum transfer, which is a function of $E_i$ and $\vartheta_f$.
This feature is commonly used to determine $\vartheta_{\rm min}$ for a given $E_i$ by assuming a fixed momentum transfer. In order to test this assertion,
we have included in Fig.6a $\vartheta_{\rm min}$ as obtained from the fixed value $\sqrt{-q^2}=1.8154$ fm$^{-1}$, corresponding to the momentum transfer at 300.5 MeV.
The deviation of this approximation from the exact phase-shift result amounts up to 1\% for $\vartheta_{\rm min}$ (Fig.6a) and up to 8\% for the corresponding cross sections  (Fig.6b)
in the energy region under consideration.
The validity of the $q$-scaling for the position of the cross section minima was also investigated experimentally \cite{Reu82}, necessitating, however, shifts in the corresponding intensities.

Fig.7a displays the energy distribution of the differential cross section at the angle of $61^\circ$,
and the experimental data from Reuter et al \cite{Reu82} and Offermann et al \cite{Off91}, spline-interpolated to $61^\circ$, are well described by the recoil-modified phase-shift analysis.
Identifying $\omega_0$ with the detector resolution from \cite{Jef20}, $\Delta E/E=5 \times 10^{-4}$, the QED- and dispersion-corrected cross section, calculated from (\ref{2.29}), is lower by about 35\%.
If the Born-type theory is used instead, the corrections are slightly smaller but tend to zero in the diffraction minimum.
A magnification of the minimum region is shown in Fig.7b, and it is seen that the predictions of (\ref{2.29}) overestimate the Jefferson Lab datum point by as much as 28\%
(which increases to 30\% if vacuum polarization is treated exactly according to (\ref{3.9})).
Were the experimental point shifted to the minimum at $60^\circ$ (keeping its cross section value unchanged), the overprediction would reduce to 3\% (respectively 5\%).
Included in Fig.7b are the respective results when the charge density provided by Offermann et al \cite{Off91} is used.
The modifications are small, about 5\% in the minimum region.

Fig.7c isolates the dispersion correction at $61^\circ$. Comparison is made between $\Delta \sigma_{\rm box}$ from the Coulomb distortion theory, calculated with two different charge densities
or with the harmonic-oscillator based $F_L$ in $C^{00}$ replaced by the exact form factor.
The use of $\varrho_N$ from \cite{Off91} deepens the minimum and reduces the maximum, while the result due to the exact $F_L$ leads to some damping of the structures.
Included is also a colculation from the Born-type theory.
However, all these models do not substantially enhance the dispersion effect and can therefore not account for the discrepancy between theory and experiment in the minimum region.

The total cross section changes $\Delta \sigma_{\rm tot}$ and $\Delta \sigma_{\rm ex}$ are displayed in Fig.7d, together with the modified result for $\Delta \sigma_{\rm ex}$ when $F_L$ is treated exactly in $C^{00}$.
While the differences between these models are small, a considerable change is introduced if the spectrometer resolution $\Delta E/E$ (affecting $\omega_0$) is reduced from $5 \times 10^{-4}$ to $2 \times 10^{-4}$.
This lowers the cross section at 362 MeV by 15\%.
An agreement between the present theory and experiment would, however, require that $\Delta E/E \approx 4 \times 10^{-5}$.

\section{Conclusion}

%Fig.7
\begin{figure*}[t]
\centering
\begin{tabular}{cc}
\hspace{-1cm}\includegraphics[width=.7\textwidth]{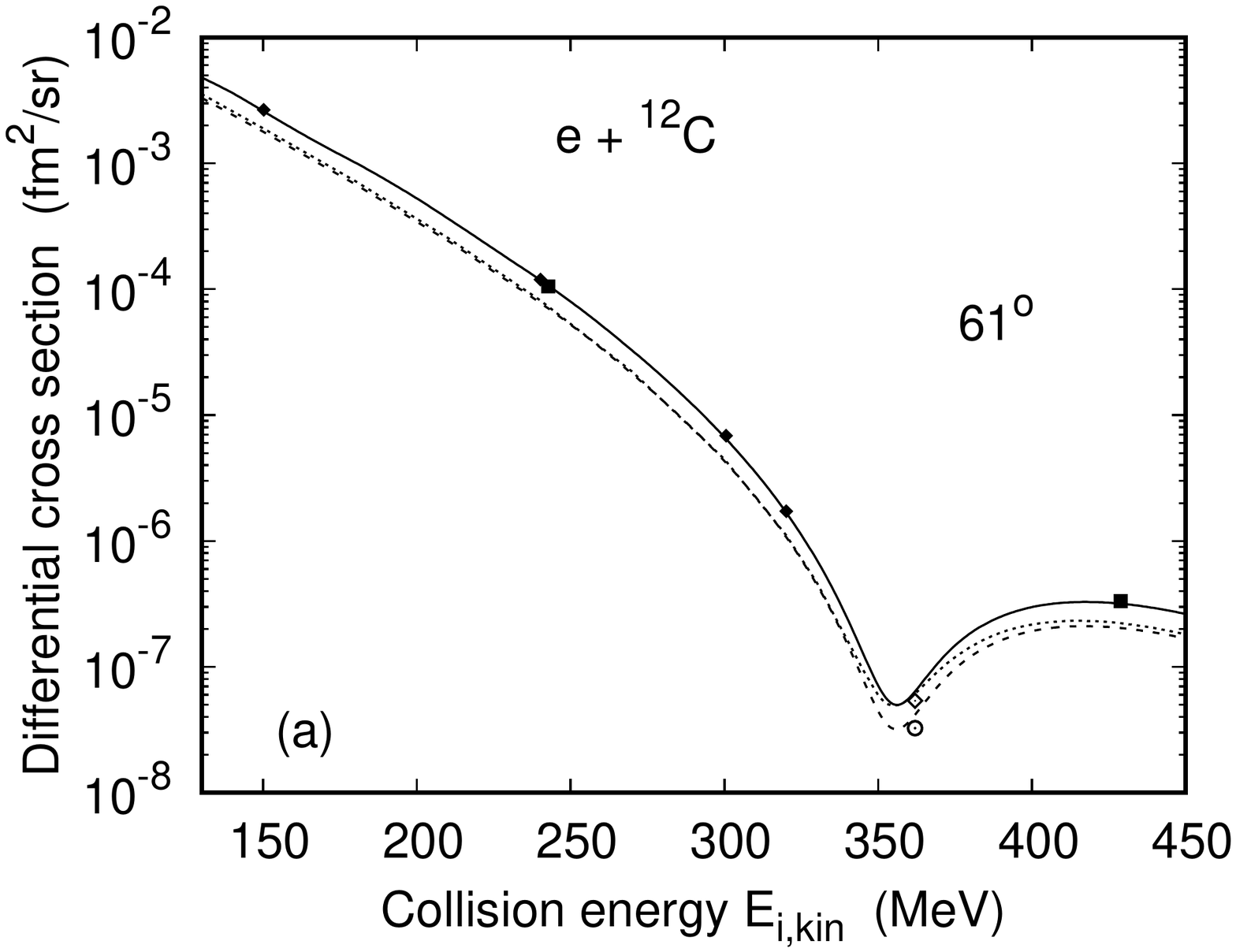}&
\hspace{-3.0cm} \includegraphics[width=.7\textwidth]{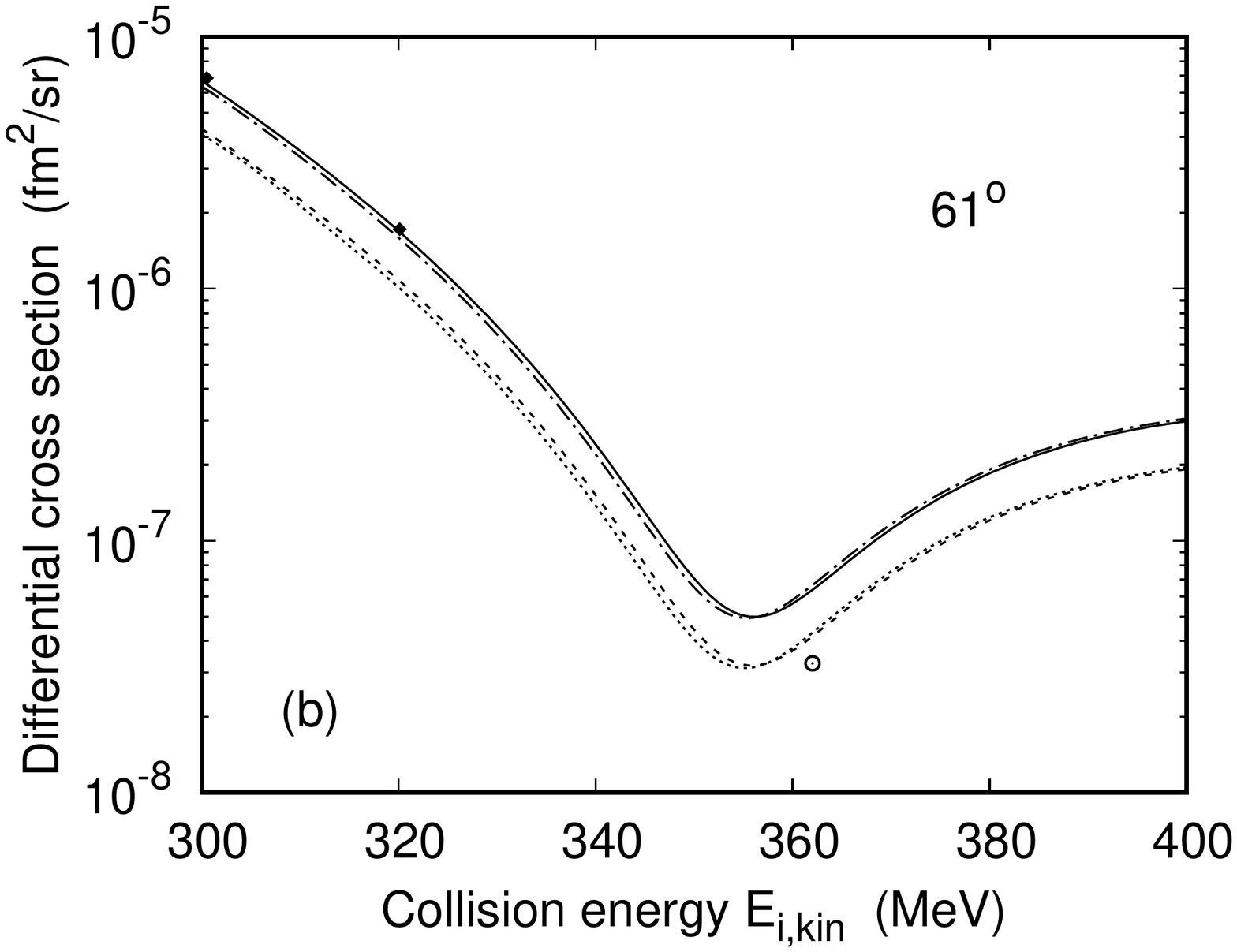}\\
\hspace{-1.cm}\includegraphics[width=.7\textwidth]{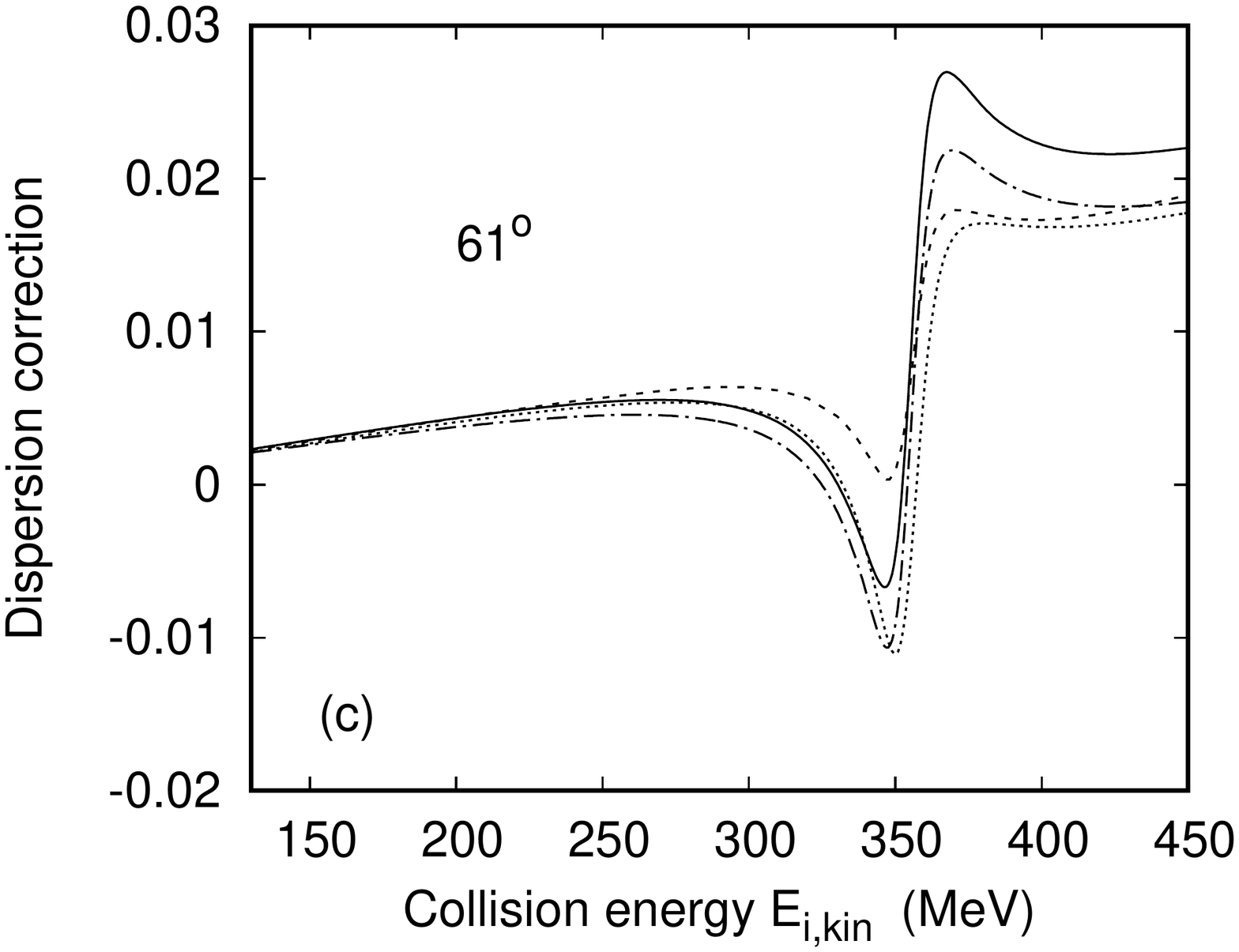}&
\hspace{-3.0cm} \includegraphics[width=.7\textwidth]{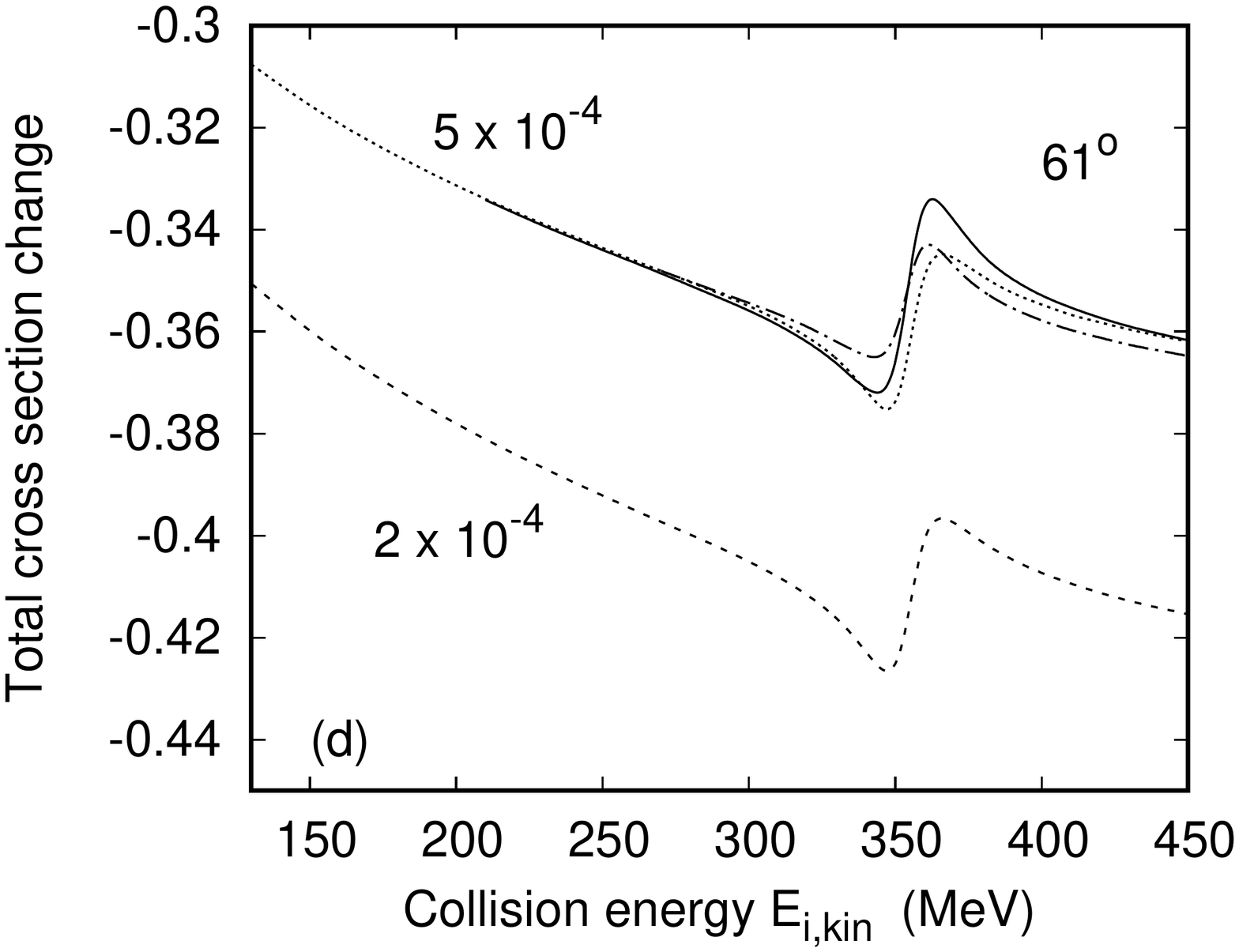}
\end{tabular}
\caption{
(a,b) Differential cross section, (c) dispersion correction and (d) total corrections for elastic electron scattering from $^{12}$C at $\vartheta_f=61^\circ$ as a function of collision energy $E_{i,{\rm kin}}$.
Shown in (a) are results from $d\sigma^{\rm coul}/d\Omega_f$ (-----------) and for the cross section including QED and dispersion effects by means of (\ref{2.29}) ($----$) and using the Born-type theory (\ref{2.27})  ($\cdots\cdots$). The experimental data are from Reuter et al ($\blacklozenge$ \cite{Reu82}), from Offermann et al ($\blacksquare$ \cite{Off91}) and from Jefferson Lab ($\odot$ \cite{Jef20}). The symbol $(\lozenge$) shows this datum point when shifted upwards by the difference between the results from (\ref{3.9}) and (\ref{2.24}), which is equal to $2.13 \times 10^{-8} $fm$^2$/sr.
The deviation of this shifted datum point from $d\sigma^{\rm coul}/d\Omega_f$ is 18\%.
In (b) the minimum region is enlarged. In addition to the results from (a) the respective results are shown when the charge density from \cite{Off91} is used instead: $d\sigma^{\rm coul}/d\Omega_f\;\, (-\cdot -\cdot -$) and total cross section from (\ref{2.29}) ($\cdots\cdots$).
Shown in (c) are $\Delta \sigma_{\rm box}$ (--------), $\Delta \sigma_{\rm box}$ but with $\varrho_N$ from \cite{Off91} ($-\cdot -\cdot -$), $\Delta \sigma_{\rm box} $ but with exact $F_L$ in $C^{00}$ ($----$) and dispersion calculated from the Born-type theory $(\cdots\cdots$).
Shown in (d) are the results for $\Delta \sigma_{\rm ex}$ (------------), $\Delta \sigma_{\rm tot} \;\,(\cdots\cdots$) and $\Delta \sigma_{\rm ex}$ but with exact $F_L$ in $C^{00}\;\, (-\cdot -\cdot -$).
Included are results for $\Delta \sigma_{\rm tot}$ if the spectrometer resolution is set to $\Delta E/E= 2\times 10^{-4}$ instead of $5 \times 10^{-4}\;\,(----)$.
}
\end{figure*}

We have estimated the QED corrections to the differential cross section for elastic scattering, resulting from vacuum polarization, from the vertex, self energy and soft bremsstrahlung correction (the vsb effect), as well as from the dispersion effect.
While vacuum polarization is treated to all orders in $Z\alpha$, the vsb effect can only be estimated in the first-order Born approximation.
The dispersion effect is obtained from the second-order Born theory with the help of a closure approximation.

The sensitivity of the vacuum polarization and the dispersion correction to various model modifications was investigated in the region of the first diffraction minimum where numerous experimental data on the $^{12}$C target are available.
Considerable differences were found, being of the order of the corrections themselves, which is in the percent region.
These model variations pass on to the total change of the cross section by the combined QED and dispersion effects.

We did not find it possible to reconcile the earlier experimental data above 200 MeV with our theory, the underprediction of the QED-corrected measurements by the theoretical cross sections (including dispersion)
amounting up to 10\% in the vicinity of the diffraction minimum, which is even higher than quoted in the literature.

There is some reason to question the conventionally used PWBA approach to the vsb correction, which contributes with $30-35\%$ the biggest portion to the cross section change.
From the investigation of vacuum polarization, both in PWBA and in an exact treatment, it follows that the exact theory exhibits structures in the diffraction region while PWBA 
predicts a monotonous behaviour with energy or angle.
It is therefore conjectured that an exact treatment of the vsb effect will show modifications near the cross section minimum which may be of considerable size since  the vsb effect itself is  very large.
This would imply an incorrect data reduction, where conventionally the vsb effect is treated as a smooth background.

A second effect which has a major influence on the vsb correction is its strong dependence on the maximum bremsstrahlung frequency $\omega_0$, which we
have identified with the experimental detector resolution.
An improvement of this resolution, respectively a decrease of $\omega_0$, will lead to an increase in magnitude of the vsb correction which, however, is only weakly dependent on energy and scattering angle.

We have also tried to interpret the recent Jefferson Lab measurement at 362 MeV, where an uncorrected experimental value of the cross section is available.
In contrast to the underprediction of the data from the other groups in the diffraction minimum by our theoretical model, the opposite is true for the 362 MeV datum point.
Also here, one might attribute the disagreement
between experiment and theory to an inaccurate account of the vsb correction.
However, it remains unclear why its behaviour should be so fundamentally different when proceeding to lower or higher collision energies.

In the literature interpretation of the 362 MeV experiment, using a PWBA-based theoretical model, it is conjectured that also in this case, the experimental cross section is above the theoretical one.
In this context it may be of importance that we cannot confirm the claim that the chosen scattering angle of $61^\circ$ leads to the minimum cross section,  taken into consideration that our recoil-corrected phase-shift approach is in this respect in perfect agreement with  the measurements at different energies.

Concerning the applicability of our model at higher impact energies, one has to cope with the problem that beyond 500 MeV the closure approximation, facilitating the dispersion estimate, will start to break down. It was shown in the context
of spin asymmetries
that for energies
in the GeV region an exact treatment of dispersion, feasible in the forward direction, leads to fundamentally higher results than obtained within the phase-shift theory. This enhancement can be traced back to the influence of highly excited
intermediate nuclear states \cite{AP04,GH08}, which are suppressed in the closure approximation. It is expected that such a dispersion behaviour will also
influence the differential cross section.

%miau

\vspace{1cm}
%\pagebreak


\begin{thebibliography}{99}

\bibitem{FW66} T.De Forest Jr. and J.D.Walecka, Adv. Phys. {\bf 15}, 1 (1966)

\bibitem{Tsa61} Y.-S.Tsai, Phys. Rev. {\bf 122}, 1898 (1961)

\bibitem{Ma69} L.C.Maximon, Rev. Mod. Phys. {\bf 41}, 193 (1969)

\bibitem{BS19} R.-D.Bucoveanus and H.Spiesberger, Eur. Phys. J. A {\bf 55}: 57 (2019)

\bibitem{MT00} L.C.Maximon and J.A.Tjon, Phys. Rev. C {\bf 62}, 054320 (2000)

\bibitem{Ub} H.\"{U}berall, {\it Electron Scattering from Complex Nuclei} (Academic Press, New York, 1971), \S3.1

\bibitem{DS84} T.W.Donnelly and I.Sick, Rev. Mod. Phys. {\bf 56}, 461 (1984)

\bibitem{AK15} A.B.Arbuzov and T.V.Kopylova, Eur. Phys. J. C {\bf 75}: 603 (2015)

\bibitem{Ueh} E.A.Uehling, Phys. Rev. {\bf 48},  55 (1935)

\bibitem{Kla77} S.Klarsfeld, Phys. Lett. {\bf 66}B, 86 (1977)

\bibitem{MU} H.Mitter and P.Urban, Acta Physica Austriaca {\bf 8}, 356 (1954)

\bibitem{L56} R.R.Lewis Jr., Phys. Rev. {\bf 102}, 544 (1956)

\bibitem{FR74} J.L.Friar and M.Rosen, Ann. Phys. {\bf 87}, 289 (1974)

\bibitem{HR98} T.Herrmann and R.Rosenfelder, Eur. Phys. J. A {\bf 2}, 29 (1998)

\bibitem{GH08} M.Gorshteyn and C.J.Horowitz, Phys. Rev. C {\bf 77}, 044606 (2008)

\bibitem{Jef20} P.Gu\`{e}ye et al (Jefferson Lab Collaboration), Eur. Phys. J .A {\bf 56}: 126 (2020)

\bibitem{Reu82} W.Reuter, G.Fricke, K.Merle and H.Miska, Phys. Rev. C {\bf 26}, 806 (1982)

\bibitem{Au11} K.Aulenbacher, Hyperfine Interact. {\bf 200}: 3 (2011)

\bibitem{Lo16} A.Lovato, S.Gandolfi, J.Carlson, S.C.Pieper and R.Schiavilla, Phys. Rev. Lett. {\bf 117}, 082501 (2016)

\bibitem{Off91} E.A.J.M.Offermann, L.S.Cardman, C.W.De Jager, H.Miska, C.De Vries and H.De Vries, Phys. Rev. C {\bf 44}, 1096 (1991)

\bibitem{Ka89} N.Kalantar-Nayestanaki et al, Phys. Rev. Lett. {\bf 63}, 2032 (1989)

\bibitem{FR76} L.W.Fullerton and G.A.Rinker Jr., Phys. Rev. A {\bf 13}, 1283 (1976)

\bibitem{BD} J.D.Bjorken and S.D.Drell, {\it Relativistic Quantum Mechanics} (McGraw-Hill, New York, 1964)

\bibitem{Gro} F.Gross, {\it Relativistic Quantum Mechanics and Field Theory} (Wiley, New York, 1993)

\bibitem{Lan} V.B.Berestetskii, E.M.Lifshitz and L.P.Pitaevskii, {\it Quantum Electrodynamics} (Course of Theoretical Physics vol.4) 2nd edition (Elsevier, Oxford, 1982), \S23, \S37, \S121.

\bibitem{Fo59} L.L.Foldy, K.W.Ford and D.R.Yennie, Phys. Rev. {\bf 113}, 1147 (1959)

\bibitem{CJ88} E.D.Cooper and B.K.Jennings, Nucl. Phys. A {\bf 483}, 601 (1988)

\bibitem{MG64} N.T.Meister and T.A.Griffy, Phys. Rev. {\bf 133}, B1032 (1964)

\bibitem{Sal} F.Salvat, J.M.Fern\'{a}ndez-Varea and  W.Williamson Jr.,  Comput. Phys. Commun.  {\bf 90}, 151 (1995)

\bibitem{YRW} D.R.Yennie, D.G.Ravenhall and  R.N.Wilson, Phys. Rev.  {\bf 95}, 500 (1954) 
 
\bibitem{VJ} H.De Vries, C.W.De Jager and  C.De Vries,  At. Data Nucl. Data Tables  {\bf 36}, 495 (1987) 


\bibitem{Jaku21} D.H.Jakubassa-Amundsen,  Eur. Phys. J.A {\bf 57}: 22 (2021)

\bibitem{Gue99} P.Gueye et al, Phys. Rev. C {\bf 60}, 044308 (1999)

\bibitem{AP04} A.V.Afanasev and N.P.Merenkov, Phys. Rev. D {\bf 70}, 073002 (2004); arXiv:0407167 [hep-ph] (2005)


\end{thebibliography}
\end{document}